\documentclass[pra,amsmath,amssymb,notitlepage,
twocolumn,superscriptaddress,longbibliography,nofootinbib,floatfix]{revtex4-2}

\usepackage{geometry}
\usepackage{tensor}
\usepackage{amsmath}
\usepackage{amssymb, mathtools}

\usepackage{tikz}
\usepackage[compat=1.0.0]{tikz-feynman}
\usepackage{enumerate}
\usepackage[hidelinks]{hyperref}

\usepackage{lipsum}
\usepackage{slashed}
\usepackage{braket}
\usepackage{yfonts}
\usepackage{graphicx}
\usepackage{dcolumn}
\usepackage{bm}

\tikzset{graviton/.style={decorate, decoration={snake, amplitude=.4mm, segment length=1.5mm, pre length=.5mm, post length=.5mm}, double}}

\begin{document}

\preprint{APS/123-QED}

\title{Might the radiation era extend back to the Big Bang? On dark matter production and the relic graviton background in quadratic gravity}

\author{Latham Boyle}\affiliation{\Edinburgh}\affiliation{\PI}
\author{Neil Turok}\affiliation{\Edinburgh}\affiliation{\PI}
\author{Vatsalya Vaibhav}\affiliation{\Edinburgh}
\newcommand*{\Edinburgh}{Higgs Centre for Theoretical Physics, James Clerk Maxwell Building, Edinburgh EH9 3FD, UK} 
\newcommand*{\PI}{Perimeter Institute for Theoretical Physics, Waterloo, Ontario N2L 2Y5, Canada}

\begin{abstract}
Naive estimates based on Einstein gravity with Lagrangian $\frac{1}{2}M_{pl}^{2}(R-2\Lambda)$ suggest that, if the radiation era extended back to the Big Bang ({\it i.e.}\ as far back as the classical spacetime background makes sense), this would result in an over-production of dark matter and a relic cosmic background of thermal gravitons. Here we revisit these conclusions in quadratic gravity, the minimal renormalizable completion of Einstein gravity, whose Lagrangian also includes the terms $\frac{1}{6}f_0^{-2}R^2-\frac{1}{2}f_{2}^{-2}C^{2}$.  Assuming the radiation era {\it does} extend back to the bang, we find a novel relation between the coefficient $f_{2}$ and the dark matter mass $m_{dm}$, needed to obtain the correct dark matter abundance.  This yields a new gravitational production mechanism for dark matter ({\it e.g.}, stable right-handed neutrinos).  Moreover, the presence or absence of the relic graviton background (detectable by forthcoming CMB experiments via its small imprint on $N_{{\rm eff}}$) will place new constraints on $f_{0}, f_{2}$.
\end{abstract}

\maketitle 


\section{Introduction}

Cosmological observations indicate that, a fraction of a second after the Big Bang, the universe was strikingly simple: the background configuration was homogeneous, isotropic, and spatially flat; with tiny ${\cal O}(10^{-5})$ random scalar perturbations that were statistically homogeneous and isotropic, adiabatic and gaussian, and with a power spectrum that was a pure power law with a nearly scale-invariant spectral index.  Moreover, these perturbations were synchronized: each spatial Fourier mode oscillated as a standing wave whose spatial phase was random, but whose temporal phase was such that, if we follow its oscillation back in time, it reached a maximum of its oscillation amplitude (with zero time derivative) at the start of the radiation era ({\it i.e.}\ it satisfied a Neumann/reflecting boundary condition at the start of the radiation era \cite{Planck:2018vyg, Mukhanov:2005sc, Weinberg:2008zzc}, as if the beginning of the radiation era was a kind of temporal mirror \cite{Boyle:2018tzc}).

It is often imagined that, in order to explain this simple early state, the radiation era only extended back to a maximal temperature well below the Planck scale ($T=T_{max}\ll M_{pl}$), and was preceded by an earlier cosmological epoch ({\it e.g.} an epoch of inflation \cite{Mukhanov:2005sc, Weinberg:2008zzc}).  But could it instead be that the radiation era extended all the way back to the Big Bang ({\it i.e.}\ as far back as the classical spacetime background makes sense),
and that the simple features we observe in the early universe reflect simple properties of the Big Bang itself \cite{Boyle:2018tzc, Turok:2022fgq, Boyle:2022lcq, Tzanavaris:2026ujl}? In this picture, those features would be explained by an appropriate quantum-gravitational theory of initial conditions, rather than being generated dynamically during an epoch of cosmic evolution between the bang and the start of the radiation era.

One concern is that (as reviewed below), if the radiation era really extended back to the bang, then when the temperature $T$ reached the Planck temperature $T\sim M_{pl}$, the gravitational scattering cross-section $\sigma$ predicted by Einstein gravity, with Einstein-Hilbert Lagrangian $\frac{1}{2}M_{pl}^{2}(R-2\Lambda)$, would imply that the dark matter should have thermalized with the radiation bath, leading to a relic dark matter abundance which is far too large. The same naive reasoning would predict a relic thermal graviton background.  

However, Einstein gravity is non-renormalizable, and the naive gravitational cross section predicted by Einstein gravity ($\sigma\sim M_{pl}^{-4}E_{cm}^{2}$, where $E_{cm}$ is the center of mass energy) is, for $E_{cm}\gtrsim M_{pl}$, incompatible with perturbative unitarity (see, {\it e.g.},~\cite{Han:2004wt}). The minimal renormalizable extension of Einstein gravity is quadratic gravity \cite{Stelle:1976gc, Salvio:2018crh}, obtained by adding to the Einstein-Hilbert Lagrangian the most general, diffeomorphism-invariant terms built from $g_{\mu\nu}$ and up to four derivatives: namely $\frac{1}{6}f_0^{-2}R^2-\frac{1}{2}f_{2}^{-2}C^{2}$ (discounting total derivative terms).  This theory initially received limited attention due to the belief that its higher-derivative structure would lead to inconsistencies, but in recent years it has seen a resurgence of interest, as a number of authors have argued that it is (or may be) consistent after all (see {\it e.g.}\ \cite{Donoghue:2021cza}).  In this context, Donoghue \& Menezes \cite{Donoghue:2018izj} showed that the gravitational cross section $\sigma$ falls as $\sigma\sim f_{2}^{4}/E_{cm}^{2}$ for $E_{cm}\gtrsim f_{2}M_{pl}$.  Hence, if $f_{2}$ $<1$, $\sigma$ remains small and is consistent with perturbative unitary at all energies.  

In this paper we reconsider the cosmology of the very early universe, assuming that quadratic gravity is correct and that the radiation era {\it does} extend back to the bang.  We find a novel relationship between the coefficient $f_{2}$ and the dark matter mass $m_{dm}$, needed to obtain the correct dark matter abundance. This is a simple new gravitational production mechanism for right-handed neutrino dark matter. (For other ideas about production of dark matter by gravity, see \cite{Chung:2001cb, Markkanen:2015xuw, Ema:2015dka,Tang:2017hvq, Bernal:2018qlk, Ema:2018ucl, Cembranos:2019qlm, Ema:2019yrd, Haro:2019umj, Haro:2019ndy, Chianese:2020yjo, Redi:2020ffc, Kolb:2020fwh, Babichev:2020yeo, Karam:2020rpa, Bernal:2020ili, Mambrini:2021zpp, Garcia:2023qab, Zhang:2023hjk, deHaro:2024asz, Racco:2024aac, Bertuzzo:2024fns, Wang:2024lva}.)  Moreover, the presence or absence of the relic graviton background (detectable by future CMB experiments via its imprint on  the effective number of neutrino species $N_{{\rm eff}}$) will yield new constraints on $f_{0}$, $f_{2}$.

\section{Background evolution}

We first review the relevant radiation-dominated FRW background solutions in quadratic gravity (QG). The action is \cite{Salvio:2018crh, Donoghue:2021cza}:
\begin{equation}
\label{higher derivative action}
  S\!=\!\int\! d^{4}x\sqrt{-g}\left[\frac{1}{2}M_{pl}^{2}R+\frac{R^{2}}{6f_0^{2}} - \frac{C^2}{2f_2^2} -\rho\right]
\end{equation}
where $R$ is the Ricci scalar, $C_{\alpha\beta\gamma\delta}$ is the Weyl curvature, $M_{pl}=(8\pi G)^{-1/2}\approx 2.4\times 10^{18}~{\rm GeV}$ is the Planck mass and $f_{0}$ and $f_{2}$ are dimensionless couplings. In describing the background, we replace the matter Lagrangian density with that for a fluid of energy density $\rho$ (including radiation, non-relativistic matter and vacuum energy).\footnote{In this paper, we regard QG as an effective field theory, and shall choose the signs of the dimensionless couplings to avoid tachyons (See Eq.~(\ref{propagator})).  We shall also ignore total derivative terms in the Lagrangian density.} For an FRW background, the line element is $ds^{2}\!=\!a^{2}(\tau)\left(-d\tau^{2}+d\Omega_{K}^{3}\right)$ (where $d\Omega_{K}^{2}$ is the line element for a maximally-symmetric 3-space of curvature $K$), $C_{\alpha\beta\gamma\delta}=0$ and $R=6\left(\frac{a''}{a^{3}}+\frac{K}{a^{2}}\right)$, where $'=d/d\tau$. Specializing to a perfect radiation fluid, $\rho=r/a(\tau)^{4}$ with $r$ constant, the solutions to Einstein-Hilbert (EH) gravity have $R=0$ (since the stress-energy tensor for radiation is traceless). The extra terms in the QG action involve the square of quantities which vanish in the EH solutions, so they make no contribution either to the equation of motion for $a(\tau)$ or to the Hamiltonian constraint\footnote{The latter is obtained by substituting $d\tau\to N (\tau) d\tau$, varying with respect to $N(\tau)$ and setting $N(\tau)=1$. For EH gravity this yields the Friedmann equation.}. Hence the usual radiation-dominated FRW solutions are also solutions of QG. The four-derivative QG equations have other solutions, like Starobinsky's original inflationary solution~\cite{Starobinsky:1980te}, 
although there has been controversy over the years (starting with \cite{Simon:1991bm}) about whether these inflating solutions, in which the higher-derivative terms dominate at early times, are really self-consistent solutions of semiclassical quantum gravity.  In any case, the backgrounds of interest to us here are those where the radiation-dominated era extends right back to the bang, and in which the higher derivative terms in the action do not effect the background evolution.

\section{Gravitational annihilation rate in quadratic gravity}

In this section we obtain the rate $\Gamma$ for gravitational annihilation of two Standard Model (thermal bath) particles into two right-handed neutrino dark matter particles, in quadratic gravity, as shown in Fig.~\ref{feynman diagram}.
The detailed calculations are presented in Appendices \ref{Propagator_Appendix}, \ref{Fermion_Graviton_Vertex}, \ref{Scalar_Graviton_Vertex}, \ref{Photon_Graviton_Vertex}.

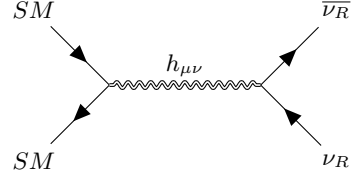
\begin{figure}[h]

    \begin{tikzpicture}  
        \begin{feynman}
  \vertex (a) at (1.5,1);
  \vertex (b) at (3.5,1); 

  \vertex (f1) at (4.5,0) {\(\nu_R\)};
  \vertex (f2) at (4.5,2) {\(\overline{\nu_R}\)};
  \vertex (i1) at (0.5,2) {\(SM\)};
  \vertex (i2) at (0.5,0) {\(SM\)};

  \diagram*{
    (i1) -- [fermion] (a) -- [fermion] (i2),
    (a) -- [graviton, edge label=\(h_{\mu\nu}\)] (b),
    (f1) -- [fermion] (b) -- [fermion] (f2),
  };
\end{feynman}
    \end{tikzpicture}
    \caption{Standard Model particles gravitationally annihilating into right-handed neutrino dark matter.}
\label{feynman diagram}
\end{figure}

For our purposes, it will be sufficient to calculate the graviton propagator in a flat background approximation, since we will see that the gravitational cross section will only be non-negligible when the wavelength of the exchanged graviton is short compared to the instantaneous Hubble length.
 
The graviton propagator is calculated in detail in Appendix \ref{Propagator_Appendix}.  (It was first calculated by Stelle in \cite{Stelle:1976gc}, but we recalculate it here using a more convenient gauge choice, and paying special attention to the $+i\epsilon$'s needed to make the Lorentzian path integral converge.)  The result is

\begin{eqnarray}
\label{propagator}
    \mathcal{D}_{\mu\nu\alpha\beta}(k)&=&\frac{4 i}{M_{pl}^{2}}\Big\{\frac{\mathcal{P}^{(2)}_{\mu\nu\alpha\beta}-\frac{1}{2}\mathcal{P}^{(0)}_{\mu\nu\alpha\beta}}{k^2}\nonumber\\ 
    &&\qquad-\frac{\mathcal{P}^{(2)}_{\mu\nu\alpha\beta}}{k^2-M_{2}^{2}}+ 
    \frac{\frac{1}{2}\mathcal{P}^{(0)}_{\mu\nu\alpha\beta}}{k^2 -M_{0}^{2}}\Big\} 
\end{eqnarray}
where $k=p+q$ is the four momentum of the graviton given as the sum of four momenta $p$ and $q$ of the incoming Standard Model particles, $M_{0}^{2}=f_{0}^{2}M_{pl}^{2}/2$ and $M_{2}^{2}=f_{2}^{2}M_{pl}^{2}/2$ define the spin-0 and spin-2 pole masses, and
\begin{eqnarray}
\label{projectors}
    \mathcal{P}^{(0)}_{\mu\nu\alpha\beta} &=& \frac{1}{3}t_{\mu\nu}t_{\alpha\beta}, \nonumber\\
    \mathcal{P}^{(2)}_{\mu\nu\alpha\beta} &=& \frac{1}{2}(t_{\mu\alpha}t_{\nu\beta}+t_{\mu\beta}t_{\nu\alpha}) - \frac{1}{3}t_{\mu\nu}t_{\alpha\beta},
\end{eqnarray}
are the corresponding spin-0 and spin-2 projectors, defined in terms of the transverse projector $t_{\mu\nu}=\eta_{\mu\nu}-k_{\mu}k_{\nu}/k^{2}$.

Note that the massive spin-0 term $\propto P^{(0)}/(k^{2}-M_{0}^{2})$ in the propagator comes from the $R^2$ term in the action, while the massive spin-2 term $\propto -P^{(2)}/(k^{2}-M_{2}^{2})$ in the propagator (which comes from the $C^2$ term in the action) has a minus (``wrong") sign: this is a ``ghost" particle, but since it is unstable and decays with lifetime of order the Planck time, it does not lead to violations of unitarity or causality on macroscopic scales. It does, however, lead to violations of {\it micro-causality}, {\it i.e.}, retrocausal effects, on Planckian scales~\cite{Lee:1969fy, Lee:1970iw, Grinstein:2008bg, Donoghue:2021meq}.

In the IR $(k^2\ll M_{0}^{2},M_{2}^{2})$, the propagator (\ref{propagator}) reduces to the propagator for Einstein gravity:
\begin{equation}
  \label{propagator_ir}
\mathcal{D}_{\mu\nu\alpha\beta}(k)=\frac{4 i}{M_{pl}^{2}}\;\frac{\mathcal{P}^{(2)}_{\mu\nu\alpha\beta}-\frac{1}{2}\mathcal{P}^{(0)}_{\mu\nu\alpha\beta}}{k^2}.
\end{equation}
So for $T\ll M_{0}^{2},M_{2}^{2}$, Fig.~\ref{feynman diagram} yields a rate $\Gamma$ of dark matter production by the thermal bath that we can estimate via standard dimensional analysis \cite{Kolb:1990vq}: the amplitude ${\cal M}\propto M_{pl}^{-2}$; so $\Gamma\propto |{\cal M}|^{2}\propto M_{pl}^{-4}$; and finally (with the power of $T$ needed to obtain the correct dimensions for $\Gamma$) we obtain the estimate $\Gamma\sim M_{pl}^{-4}T^{5}$.

In the UV $(k^2\gg M_{0}^{2},M_{2}^{2})$, the propagator (\ref{propagator}) reduces to\footnote{In the present paper, the RG flow of $f_{0}$ and $f_{2}$ may be neglected; but this flow is studied in \cite{Boyle:2025bxf}, where we highlight an intriguing Standard-Model-like theory where the gravitational couplings have fixed points.}
\begin{equation}
  \label{propagator_uv}
  {\cal D}_{\mu\nu\alpha\beta}=-\frac{2if_{2}^{2}{\cal P}^{(2)}_{\mu\nu\alpha\beta}}{k^{4}}
  +\frac{if_{0}^{2}{\cal P}^{(0)}_{\mu\nu\alpha\beta}}{k^{4}}
\end{equation}
So for $T\gg M_{0}^{2}$, $M_{2}^{2}$, we repeat the preceding analysis: now, since the contribution from the spin-0 part of the graviton propagator may be neglected\footnote{This is because the vertex rule for matter-antimatter-(scalar)graviton vertex is proportional the trace of the stress-tensor of the matter field, which is either zero (for conformally-coupled fields) or proportional to the its mass, which is negligible compared to the Planckian energies of interest, see Appendices \ref{Fermion_Graviton_Vertex}, \ref{Scalar_Graviton_Vertex}, \ref{Photon_Graviton_Vertex}.} we have ${\cal M}\propto f_{2}^{2}$; so $\Gamma\propto|{\cal M}|^{2}\propto f_{2}^{4}$; and finally (with the power of $T$ needed to obtain the right dimensions) we obtain the estimate $\Gamma\sim f_{2}^{4}T$.

In Appendices \ref{Fermion_Graviton_Vertex}, \ref{Scalar_Graviton_Vertex}, \ref{Photon_Graviton_Vertex}, \ref{App:Gamma_Bar}, we calculate in detail the 
average rate $\overline{\Gamma}$ for two particles in the thermal bath to gravitationally annihilate into two dark matter particles (via the process shown in Fig.~\ref{feynman diagram}),
obtaining the result (see Eqs.~(\ref{Gamma_bar_Appendix}, \ref{C_Appendix}))
\begin{eqnarray}
\label{Gamma(T)}
  \overline{\Gamma}(T) = \left\{\begin{array}{ll} 
  C_{\,{\rm IR}\,}\,M_{pl}^{-4}\,T^5 & \quad(T< T_{\ast}) \\
  C_{{\rm UV}}\,f_2^4\,T & \quad(T> T_{\ast})\end{array}\right.
\end{eqnarray}
where 
\begin{subequations}
\label{C_Appendix}
  \begin{eqnarray}
      C_{\,{\rm IR}\;}&=&\;\;\,\frac{1393}{25216 \pi^{3}}\;\;\,\approx
      2.1\times 10^{-3}, \\
      C_{{\rm UV}}&=&\frac{1393}{19365888 \pi^3}\approx
      2.8\times 10^{-6}.
  \end{eqnarray}
\end{subequations}
Equating these two rates, we find the transition temperature 
\begin{equation}
  \label{T_star}
  T_{\ast}= \left(C_{{\rm UV}}/C_{{\rm IR}}\right)^{1/4}f_2 M_{pl}
  =\frac{f_{2}M_{pl}}{3^{1/4}\times 4}.
\end{equation}

\section{Gravitational dark matter production in quadratic gravity}

In this section, we use Eq.~(\ref{Gamma(T)}) for $\overline{\Gamma}$ to derive the relic dark matter abundance, and derive constraints on the quadratic gravity coupling $f_{2}$ and dark matter mass $m_{dm}$.

Since we are assuming the standard (inflation-free) radiation-era evolution, in which spatial curvature is unimportant at early times, we may use the standard Friedmann equation $H^{2}=\rho/(3 M_{pl}^{2})$, which becomes
\begin{equation}
  \label{H(T)}
  H=\left(\frac{\pi^{2}}{90}g_{\ast}\right)^{1/2}
  \frac{T^{2}}{M_{pl}}
\end{equation}
where we have used $\rho=(\pi^{2}/30)g_{\ast}T^{4}$ for the energy density of a relativistic thermal bath \cite{Kolb:1990vq}, and $g_{\ast}=110.25$ is the effective number of species in the high-temperature thermal bath, with fermion species are weighted by $7/8$ (see Eqs. (3.61, 3.62) in Ref.~\cite{Kolb:1990vq} and Eq.~(\ref{gstar_sm}) below).

\begin{figure}
    \centering
    \includegraphics[width=1\linewidth]{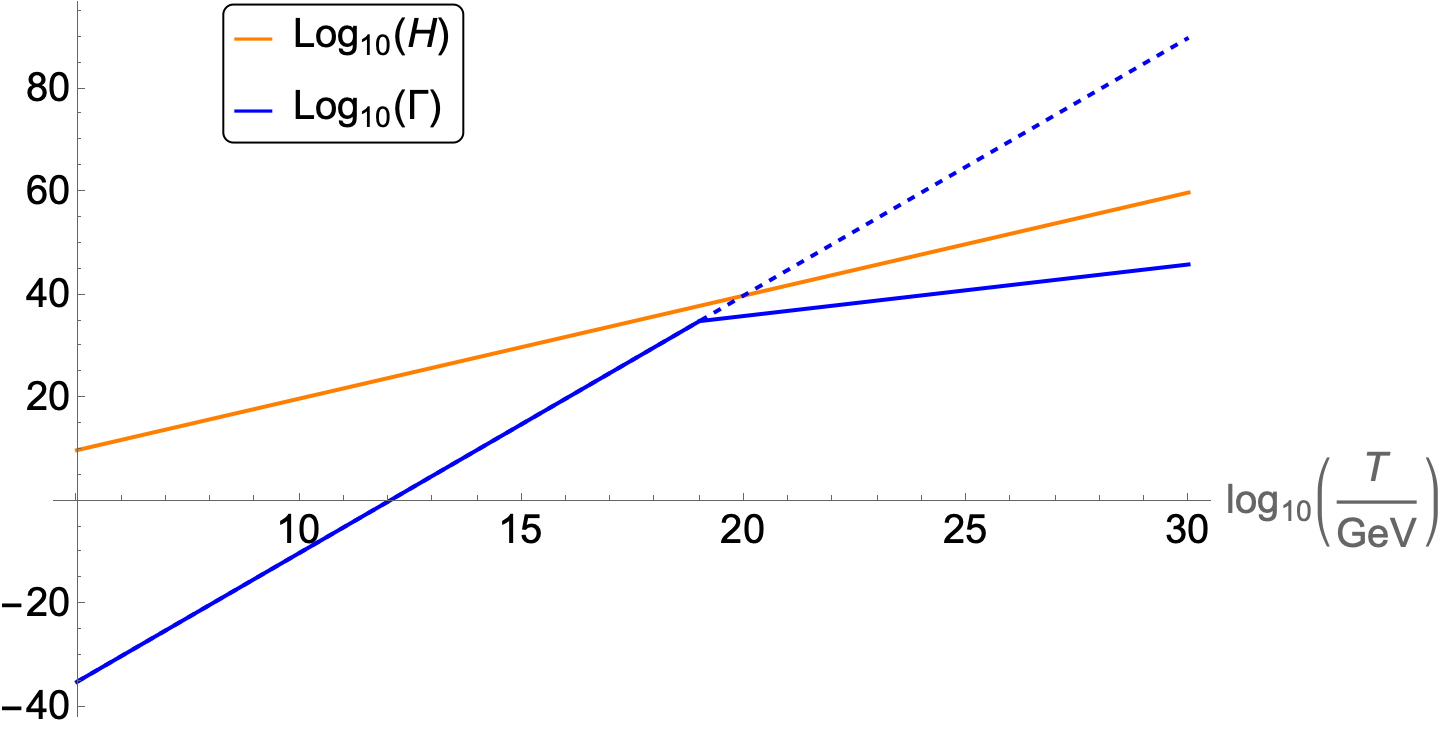}
    \caption{Comparison of the gravitational annihilation rate $\Gamma(T)$ with the Hubble rate $H(T)$.  In Einstein gravity (dashed blue line), $\Gamma(T)>H(T)$ (and unitarity is violated) when $T\gtrsim M_{pl}$.  In quadratic gravity (solid blue line), the unitarity bound $f_{2}\lesssim{\cal O}(1)$ implies $\Gamma(T)\lesssim H(T)$ for all $T$, and $f_{2}\ll 1$ implies $\Gamma(T)\ll H(T)$, so that dark matter never thermalizes.}
    \label{dark matter plot}
\end{figure}

The unitarity bound \cite{Han:2004wt} implies that the steep growth of $\Gamma(T)\propto T^5$ predicted by Einstein gravity cannot extend beyond the Planck scale, so we must have $T_{\ast}\lesssim M_{pl}\Rightarrow f_{2}<{\cal O}(1)$.  This, in turn, is roughly the same as the condition that the dark matter never equilibrated with the radiation bath (corresponding to the condition that the ratio $\overline{\Gamma}(T)/H(T)$ was $<1$ for all $T$ \cite{Kolb:1990vq} since, as may be seen from Fig.~\ref{dark matter plot}, this ratio was largest at $T=T_{\ast}$, and the conditions $T_{\ast}\lesssim M_{pl}$ and $\overline{\Gamma}(T_{\ast})\lesssim H(T_{\ast})$ yield nearly the same bound on $f_{2}$).  But a different (stronger) constraint on $f_{2}$ follows from avoiding dark matter overproduction, as we now explain.

In Appendix \ref{App:Gamma_Bar}, we show that, just after the era of dark matter production by annihilation of thermal bath particles has ended, the number of 
dark matter particles per thermal bath particle is given by (see Eq.~\ref{n_dm_over_n_tot})
\begin{equation}
  \label{n_dm_over_n_tot_body}
    \frac{n_{dm}}{n_{tot}}
    =2\int \overline{\Gamma}(t')dt'.
\end{equation}
Substituting Eq.~(\ref{Gamma(T)}) into 
Eq.~(\ref{n_dm_over_n_tot_body}) and using the fact that, when $\overline{\Gamma}(T)\propto T^{n}$ in the radiation era, we have
$\int_{0}^{t_{\ast}\,}\overline{\Gamma}(t')dt'
  =\frac{1}{2-n}\left.\frac{\overline{\Gamma}}{H}\right|_{t_{\ast}}$ (for $n<2$) or 
  $\int_{t_{\ast}}^{\infty}\overline{\Gamma}(t')dt'
  =\frac{1}{n-2}\left.\frac{\overline{\Gamma}}{H}\right|_{t_{\ast}}$ (for $n>2$),

(\ref{n_dm_over_n_tot_body}) becomes
\begin{equation}
    \frac{n_{dm}}{n_{tot}}=
    2(\frac{1}{1}\!+\!\frac{1}{3})
    \!\left.\frac{\overline{\Gamma}}{H}\right|_{T_{\ast}}
    \!\!=\!
    \frac{1393(300)^{1/4}}
    {605184\pi^{4}}\frac{f_{2}^{3}}{g_{\ast}^{1/2}}.
\end{equation}

\begin{figure}[h]
    \centering
    \includegraphics[width=0.9\linewidth]{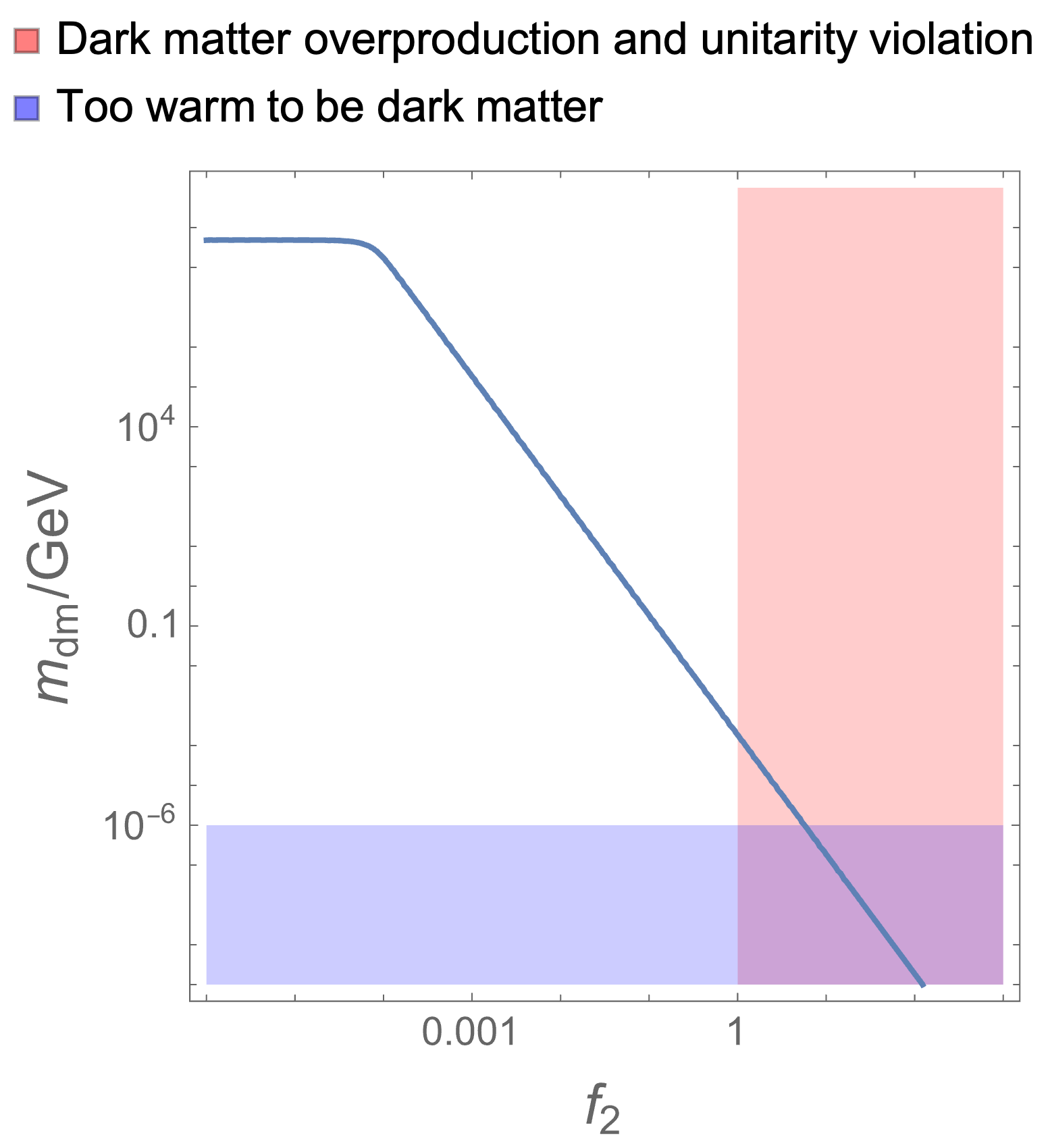}
    \caption{Constraint on the Weyl coupling $f_2$ and the dark matter mass $m_{dm}$ coming from conservation of dark matter yield.}
    \label{f2_m_dm_plot}
\end{figure}

From here we can obtain an expression for the 
quantity $n_{dm}/s$, which is conserved between the early radiation era (just after the dark matter finished being produced) and the present, by writing $n_{dm}/s=(n_{dm}/n_{tot})(n_{tot}/s)$, and using the fact that $s=g_{\ast} \frac{2\pi^{2}}{45}T^{3}$ and $n_{tot}=g_{\ast n}\frac{\zeta(3)}{\pi^{2}}T^{3}$, where $g_{\ast n}=98.5$ is the effective number of species in the high-temperature thermal bath, with fermion species are weighted by $3/4$ (see Eqs. (3.52) in Ref.~\cite{Kolb:1990vq} and Eq.~(\ref{gstar_n_sm}) below), to obtain
\begin{eqnarray}
  \!\!\!\!\!\!\!\!\!\!\!\frac{n_{dm}}{s}\!=\!
  \frac{45\zeta(3)}{2\pi^{4}}\frac{\!1393(300)^{\frac{1}{4}}\!}
    {605184\pi^{4}}\frac{\!g_{\ast n}f_{2}^{3}\!}{g_{\ast}^{3/2}}\approx 2.3\!\times\! 10^{-6}\!f_{2}^{3}.
\end{eqnarray}
Note that, if the universe is in its CPT-symmetric vacuum state, dark matter is also gravitationally produced via the cosmological Hawking-Unruh effect \cite{Boyle:2018rgh, Boyle:2018tzc}.  Including this additional contribution (calculated in \cite{Boyle:2018rgh, Boyle:2018tzc}), $n_{dm}/s$ becomes
\begin{equation}
  \label{n_dm_over_s}
    \!\!\frac{n_{dm}\!}{s} \!= 
    8.1\!\times\!10^{-32}
    \!\left(\frac{m_{dm}}{{\rm GeV}}\right)^{3/2}\!\!+2.3\times 10^{-6}f_{2}^{3}.
\end{equation}
Assuming the standard adiabatic thermal evolution between then and now (since no first-order phase transition is expected in the Standard Model), $n_{dm}/s$ is conserved and is equal to its present value $n_{dm,0}/s_{0}$, which we can write as (again, see Section 3.4 in \cite{Kolb:1990vq})
\begin{eqnarray}
  \label{n_dm_over_s_0}
  \frac{n_{dm,0}}{s_{0}}  
  &=&\frac{1}{7.04}\frac{n_{dm,0}}
  {n_{b,0}}\frac{n_{b,0}}{n_{\gamma,0}} =\frac{1}{7.04}\frac{m_{p}}{m_{dm}} \frac{\Omega_{dm,0}}{\Omega_{b,0}}
  \frac{n_{b,0}}{n_{\gamma,0}} \nonumber\\  &\approx&4.3\times10^{-10}\left(\frac{m_{dm}}{{\rm GeV}}\right)^{-1}.
\end{eqnarray}
where, in the final expression, we have substituted the measured values \cite{Planck:2018vyg} 
$\Omega_{dm,0}h^{2}=0.120\pm0.001$, $\Omega_{b,0}h^{2}=0.0224\pm0.0001$, 
and (see p 78 in \cite{Kolb:1990vq}) $\eta=n_{b,0}/n_{\gamma,0}=2.68\times 10^{-8}\Omega_{b,0}h^{2}$.

\begin{figure}[h]
\resizebox{3.5 cm}{!}{
\begin{tikzpicture}
  \begin{feynman}
  \vertex (a) at (1,1){$\gamma$};
  \vertex (b) at (4,1){$h_{\mu\nu}$};
  \vertex (c) at (1,4){$\gamma$};
  \vertex (d) at (4,4){$h_{\mu\nu}$};
  \vertex (v) at (2.5,2.5);
    \diagram*{
      (a)[particle=\(\gamma\)] -- [photon] (v),
      (b)[particle=\(h_{\mu\nu}\)] -- [graviton] (v),
      (c)[particle=\(\gamma\)] -- [photon] (v),
      (d)[particle=\(h_{\alpha\beta}\)] -- [graviton] (v),
      (v) [dot],
    };
  \end{feynman}
\end{tikzpicture}
}
\caption{The two-gauge-boson/two-graviton vertex.} 
\label{two_photons_two_gravitons}
\end{figure}
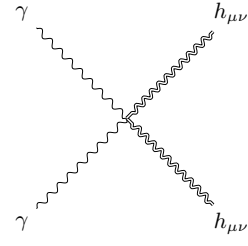

Equating expressions (\ref{n_dm_over_s}) and (\ref{n_dm_over_s_0}) for $n_{dm}/s$, yields the relationship between $f_{2}$ and $m_{dm}$ plotted Fig.~\ref{f2_m_dm_plot}.  We see that
that the gravitational production of dark matter in the early universe breaks into two regimes: (i) for $f_2 \lesssim 7\times 10^{-5}$, it proceeds via the later non-thermal (cosmological Hawking-Unruh) mechanism discussed in \cite{Boyle:2018tzc, Boyle:2018rgh}, so that the previous prediction $m_{dm}\!\approx\! 5\!\times\! 10^{8}{\rm GeV}$ \cite{Boyle:2018tzc, Boyle:2018rgh} is recovered; (ii) for $f_{2} \gtrsim 7\times 10^{-5}$, it proceeds via the earlier thermal mechanism calculated in this paper, with lower dark matter masses $m_{dm}\approx 2\times10^{-4}f_{2}^{-3}~{\rm GeV}$ predicted.

\section{Graviton non-thermalization}

In Einstein gravity, if the radiation era extends back to $T\sim M_{pl}$, gravitons equilibrate with the thermal bath.  What about in quadratic gravity?  To answer this, we must check whether the relevant thermalizing processes (see Fig.~\ref{two_photons_two_gravitons} for an example process) ever had rate $\Gamma(T)>H(T)$ \cite{Kolb:1990vq}, just as we did above for dark matter.

For this question, it is convenient to switch to the convention where the graviton's kinetic term is canonically normalized (by defining $h_{\mu\nu}^{\rm IR}=M_{pl}h_{\mu\nu}$ in the IR and $h_{\mu\nu}^{\rm UV}=f_{2}^{-1}h_{\mu\nu}$ in the UV).  In this convention,
we see every $h_{\mu\nu}$ leg on every vertex in a Feynman diagram contributes a factor of $M_{pl}^{-1}$ in the IR and $f_{2}$ in the UV,
so that at leading order any process has rate $\Gamma_{IR}(T)\sim M_{pl}^{-2n}T^{2n+1}$ and $\Gamma_{UV}(T)\sim f_{2}^{2n}T$, with $n\geq 1$ and transition temperature $T_{\ast}\sim f_{2}M_{pl}$.  And from Eq.~(\ref{H(T)}) we have $H(T)\sim M_{pl}^{-1}T^{2}$.  We see that $\Gamma(T)<H(T)$ for all $T$ if $\Gamma(T_{\ast})<H(T_{\ast})$ or, equivalently, $f_{2}\lesssim 1$.  A similar calculation shows that, to avoid graviton thermalization at high $T$, we also need $f_{0}\lesssim 1$.

Stated another way: if $f_{0}$ and $f_{2}$ are both $\ll1$, gravitons did not thermalize, but if either $f_{0}$ or $f_{2}$ is $\sim 1$, they did.  We can test which possibility occurred: if the gravitons thermalized, the universe today should contain a relic cosmic graviton background (along with the cosmic microwave and neutrino backgrounds).  As shown in Appendix \ref{N_eff}, if such a graviton background exists, it causes a shift in the number of effective neutrino species $N_{eff}$ from its standard value $N_{eff}=3.046$ to a larger value $N_{eff}=3.046+0.051=3.097$.  Both predictions are compatible with current observations \cite{ParticleDataGroup:2024cfk, Planck:2018vyg, DiValentino:2019dzu}, but may be distinguished by future experiments including the Simons Observatory \cite{SimonsObservatory:2018koc} (see Appendix \ref{N_eff} for details).

\section{Acknowledgements}
We thank Niayesh Afshordi, Sam Bateman, Brian Batell, John Donoghue, Gordan Krnjaic and Maxim Pospelov for helpful discussions. VV is supported by the School of Physics and Astronomy Studentship at the University of Edinburgh. LB and NT are supported by STFC Consolidated Grant ‘Particle Physics at the Higgs Centre,’ and  NT is supported by the Higgs Chair at the University of Edinburgh. Perimeter Institute is supported by the Government of Canada, via Innovation, Science and Economic Development, Canada and by the Province of Ontario via the Ministry of Research, Innovation and Science.

\bibliography{references}

\appendix

\section{Quadratic gravity propagator}
\label{Propagator_Appendix}
In this appendix, we carefully derive the propagator for quadratic gravity, for a general gauge parameter $\xi$, and also in a particularly convenient gauge, paying careful attention to signs, and the various $+i\epsilon$ prescriptions needed to make the oscillatory Lorentzian path integral converge.

\subsection{Useful expansions in $h_{\mu\nu}$}
We will use the ``$+++$" conventions of Misner, Thorne, and Wheeler \cite{Misner:1973prb} (shared {\it e.g.}\ by Wald \cite{Wald:1984rg} and Carroll \cite{Carroll:2004st}), so the metric signature is ``mostly plus": $(-,+,+,+)$; and the curvature tensors are defined as
\begin{subequations}
\begin{eqnarray}
  R^{\alpha}_{\;\;\beta\gamma\delta}&=&
  \Gamma^{\alpha}_{\beta\delta,\gamma}
  -\Gamma^{\alpha}_{\beta\gamma,\delta}
  +\Gamma^{\alpha}_{\sigma\gamma}\Gamma^{\sigma}_{\beta\delta}
  -\Gamma^{\alpha}_{\sigma\delta}\Gamma^{\sigma}_{\beta\gamma}\qquad \\
  R_{\beta\delta}&=&R^{\alpha}_{\;\;\beta\alpha\delta} \\
  R&=&g^{\alpha\beta}R_{\alpha\beta}
\end{eqnarray}
\end{subequations}
where the Christoffel symbol $\Gamma^{\alpha}_{\beta\gamma}$ is
given by
\begin{equation}
\Gamma^{\alpha}_{\beta\gamma}=
    \frac{1}{2}g^{\alpha\delta}(g_{\beta\delta,\gamma}
    +g_{\gamma\delta,\beta}-g_{\beta\gamma,\delta}). 
\end{equation}
Now we write the perturbed metric as
\begin{equation}
    g_{\mu\nu}=\eta_{\mu\nu}+h_{\mu\nu}
\end{equation}
so the inverse metric, up to ${\cal O}(h^{2})$, is
\begin{equation}
    g^{\mu\nu}=\eta^{\mu\nu}-h^{\mu\nu}+h^{\mu\rho}h_{\rho}^{\;\;\nu}+{\cal O}(h^3), 
\end{equation}
and $\sqrt{-g}$, up to ${\cal O}(h)$, is
\begin{equation}
  \sqrt{-g}=1+\frac{1}{2}h+{\cal O}(h^2)
\end{equation}
where, indices on $h_{\mu\nu}$ and $\partial_{\mu}$ are raised and lowered using $\eta_{\mu\nu}$, and 
\begin{equation}
    h\equiv h^{\alpha}_{\alpha}.
\end{equation}
Then we obtain the following useful intermediate expansions
\begin{subequations}
\begin{eqnarray}
    \Gamma^{\alpha}_{\beta\gamma}
    &=&\frac{1}{2}(h^{\alpha}_{\beta,\gamma}+h^{\alpha}_{\gamma,\beta}-h_{\beta\gamma}^{\;\;\;\;,\alpha}) \nonumber\\
    &-&\frac{1}{2}h^{\alpha\delta}
    (h_{\beta\delta,\gamma}+h_{\gamma\delta,\beta}-h_{\beta\gamma,\delta}) \nonumber\\
    &+&{\cal O}(h^3) \\
    R&=&{h^{\alpha\beta}}_{,\alpha\beta}-\Box h \nonumber\\
    &+&h^{\alpha\beta}(\Box h_{\alpha\beta}+h_{,\alpha\beta}
    -2h^{\gamma}_{\alpha,\beta\gamma}) \nonumber\\
    &-&h^{\gamma,\alpha}_{\alpha}h_{\gamma,\beta}^{\beta}
    +h^{\gamma,\alpha}_{\alpha}h_{,\gamma}
    -\frac{1}{4}h_{,\gamma}h^{,\gamma} \nonumber\\
    &+&\frac{3}{4}h^{\alpha\beta,\gamma}h_{\alpha\beta,\gamma}
    \!-\!\frac{1}{2}h_{\alpha\beta,\gamma}h^{\beta\gamma,\alpha}
    \nonumber\\
    &+&{\cal O}(h^3)\\
    R_{\alpha\beta}&=&\frac{1}{2}(h^{\gamma}_{\alpha,\beta\gamma}+h^{\gamma}_{\beta,\alpha\gamma}-\Box h_{\alpha\beta}-h_{,\alpha\beta})\nonumber\\
    &+&{\cal O}(h^2).
\end{eqnarray}
\end{subequations}
Using the above expressions we obtain (after integrating by parts and dropping total derivative terms), the following expressions for the various relevant terms in the action (up to quadratic order in $h$):
\begin{subequations}
\begin{eqnarray}
  &&\int d^{4}x\sqrt{-g}R=\int d^{4}x\;\frac{1}{4}\Big[h^{\alpha\beta}\Box h_{\alpha\beta}
  -h\Box h \nonumber\\
  &&+2h^{\alpha\beta}h_{,\alpha\beta} 
  +2h^{\alpha}_{\gamma,\alpha}h_{\beta}^{\gamma,\beta}\Big] \\
  &&\int d^{4}x\sqrt{-g}R^{2}=
  \int d^{4}x({h^{\alpha\beta}}_{,\alpha\beta}-\Box h)^{2} \\
  &&\int d^{4}x \sqrt{-g}R_{\alpha\beta}^{2}=
  \int d^{4}x\;\frac{1}{4}\Big[(\Box h_{\alpha\beta})^{2}
  +(\Box h)^{2} \nonumber\\
  &&-2{h^{\alpha\beta}}_{,\alpha\beta}\Box h+
  2({h^{\alpha\beta}}_{,\alpha\beta})^{2}
  +2h^{\alpha}_{\gamma,\alpha}\Box h^{\gamma,\beta}_{\beta}\Big] \\
  &&\int d^{4}x \sqrt{-g}C^{2}=
  \int d^{4}x\Big[\frac{1}{2}(\Box h_{\alpha\beta})^{2}\!-\!\frac{1}{6}(\Box h)^{2} \nonumber\\
  &&+\frac{1}{3}{h^{\alpha\beta}}_{,\alpha\beta}\Box h+\frac{1}{3}({h^{\alpha\beta}}_{,\alpha\beta})^{2}+h^{\alpha}_{\gamma,\alpha}\Box h_{\beta}^{\gamma,\beta}\Big].\qquad
\end{eqnarray}
\end{subequations}
In the final expression for $C^2$, we have used 
\begin{equation}
    C^{2}=
    R_{\alpha\beta\gamma\delta}^{2}-2R_{\alpha\beta}^{2}
    +\frac{1}{3}R^{2} 
    =(2R_{\alpha\beta}^{2}-\frac{2}{3}R^{2})+E
\end{equation}
where $E=R_{\alpha\beta\gamma\delta}^{2}-4R_{\alpha\beta}^{2}+R^{2}$ is the 4D Euler density (the Gauss-Bonnet curvature), and $\sqrt{-g}E$ is a total derivative which may be discarded.

\subsection{Deriving propagators}

In this subsection, we derive the propagators for quadratic gravity in a convenient gauge, first reviewing the Klein-Gordon, Maxwell and Einstein-Hilbert cases to warm up and introduce notation.

\subsubsection{Klein-Gordon}

First consider the Klein-Gordon (``KG'') action
\begin{equation}
  S_{KG}=\int d^{4}x\frac{1}{2}\varphi {\cal O}\varphi
\end{equation}
with
\begin{equation}
  {\cal O}=\partial^{2}-m^{2}+i\epsilon
\end{equation}
where we have made the substitution ${\cal O}\to{\cal O}+i\epsilon$ to make the oscillatory path integral $\int {\cal D}\varphi {\rm e}^{iS_{KG}[\varphi]}$ converge.  

Next we go to momentum space
\begin{equation}
  {\cal O}(\partial_{\mu}\to i k_{\mu})
  =-k^{2}-m^{2}+i\epsilon.
\end{equation}
The propagator is then the operator ${\cal D}$ satisfying
\begin{equation}
  \label{KG_propagator_definition}
  {\cal O}{\cal D}=i
\end{equation}
yielding the usual Klein-Gordon propagator:
\begin{equation} 
  \label{KG_propagator_prelim}
    {\cal D}_{KG}=\frac{i}{-k^{2}-m^{2}+i\epsilon}.
\end{equation}

\subsubsection{Maxwell}

Next consider the Maxwell action
\begin{eqnarray}
  S_{M}=\int d^{4}x\left[-\frac{1}{4}F_{\mu\nu}^{2}\right].
\end{eqnarray}
In this case, because the action is gauge-invariant
under $A_{\mu}\to A_{\mu}+\partial_{\mu}\alpha$, we must add a gauge-fixing term before we can invert the kinetic operator.\footnote{And really we should also add a ghost term, but this is not needed for determining the photon propagator, so we ignore it here.}  With the usual Lorentz gauge-fixing condition 
\begin{equation}
  f(A)=\partial_{\mu}A^{\mu}=0
\end{equation}
the Fadeev-Popov procedure leads us to add the gauge-fixing 
term $S_{GF}=-\frac{1}{2\xi}\int d^{4}x\,f(A)^{2}$, so 
\begin{eqnarray}
  S&=&\int d^{4}x\left[-\frac{1}{4}F_{\mu\nu}^{2}
  -\frac{1}{2\xi}f(A)^{2}\right] \\
  &=&\int d^{4}x \frac{1}{2}A^{\mu}{\cal O}_{\mu\nu}A^{\nu}
\end{eqnarray}
with 
\begin{equation}
  {\cal O}_{\mu\nu}=\eta_{\mu\nu}\Box+\left(\frac{1}{\xi}-1\right)\partial_{\mu}\partial_{\nu}+i\epsilon \eta_{\mu\nu}
\end{equation}
where we have again made the substitution ${\cal O}_{\mu\nu}\to{\cal O}_{\mu\nu}+i\epsilon\eta_{\mu\nu}$ to make the oscillatory path integral $\int {\cal D}\varphi \,{\rm e}^{iS_{M}[A_{\mu}]}$ converge.

Next we go to momentum space
\begin{equation}
  {\cal O}_{\mu\nu}(\partial_{\mu}\!\to\!i k_{\mu})
  \label{O_Maxwell_kspace_projectors}
  =(-k^{2}+i\epsilon)t_{\mu\nu}
  +(-\frac{1}{\xi}k^{2}+i\epsilon)\ell_{\mu\nu}
\end{equation}
where we have defined the transverse and longitudinal projection operators,
\begin{equation}
    t_{\mu\nu}\equiv\eta_{\mu\nu}-\frac{k_{\mu}k_{\nu}}{k^{2}},\qquad
    \ell_{\mu\nu}\equiv \frac{k_{\mu}k_{\nu}}{k^{2}},
\end{equation}
which are orthogonal,
\begin{equation}
t^{\mu}_{\;\;\rho}t^{\rho}_{\;\;\nu}=t^{\mu}_{\;\;\nu},\quad
  \ell^{\mu}_{\;\;\rho}\ell^{\rho}_{\;\;\nu}
  =\ell^{\mu}_{\;\;\nu},\quad
  t^{\mu}_{\;\;\rho}\ell^{\rho}_{\;\;\nu}=0,
\end{equation}
and sum to the identity,
\begin{equation}
  t_{\mu\nu}+\ell_{\mu\nu}=\eta_{\mu\nu}.
\end{equation}

The propagator is the operator ${\cal D}_{\mu\nu}$ satisfying
\begin{equation}
  \label{Maxwell_propagator_definition}
  {\cal O}^{\mu}_{\;\;\rho}{\cal D}^{\rho}_{\;\;\nu}=i\delta^{\mu}_{\;\;\nu}.
\end{equation}
Since $t_{\mu\nu}$ and $\ell_{\mu\nu}$ are orthogonal projectors, ${\cal D}_{\mu\nu}$ can be obtained by inverting their coefficients and multiplying by $i$:
\begin{equation}
  {\cal D}_{\mu\nu}^{M}=\frac{i}{-k^{2}+i\epsilon}t_{\mu\nu}+\frac{i\xi}{-k^{2}+i\epsilon\xi}\ell_{\mu\nu}
\end{equation}
which is equivalent to the standard result, except that, since we have been careful to use the $i\epsilon$'s associated with the shift ${\cal O}_{\mu\nu} \to {\cal O}_{\mu\nu}+i\epsilon\eta_{\mu\nu}$, making the path integral converge, we find a $+i\epsilon\xi$ in the denominator of the (gauge-dependent) longitudinal ($\ell_{\mu\nu}$) term, whereas in the standard expression, this is usually replaced by $+i\epsilon$, for simplicity.  
 
\subsubsection{Einstein-Hilbert}

Now consider the Einstein-Hilbert (``EH'') action
\begin{subequations}
  \begin{eqnarray}
    S_{{\rm EH}}&=&\int d^{4}x\sqrt{-g}\frac{1}{16\pi G}
    R \\
    &\approx& \int d^{4}x\frac{1}{2}
    \Big[h^{\alpha\beta}\Box h_{\alpha\beta}-h\Box h
    \nonumber\\
    &&\qquad\qquad+2h^{\alpha\beta}h_{,\alpha\beta}
    +2h^{\alpha}_{\gamma,\alpha}h_{\beta}^{\gamma,\beta}\Big]\qquad
  \end{eqnarray}
\end{subequations}
where, in this subsection, we have made the substitution $h_{\mu\nu}\to (32\pi G)^{1/2}h_{\mu\nu}$
to make $h_{\mu\nu}$ a dimension-one field.
In this case, because the action is gauge-invariant under $h_{\mu\nu}\to h_{\mu\nu}+\partial_{\mu}\xi_{\nu}+\partial_{\nu}\xi_{\mu}$, we again have to add a gauge-fixing term before we can invert the kinetic operator.  With the usual de Donder gauge-fixing condition ({\it i.e.},\ the analogue of the Lorentz gauge condition, but applied to the trace-reversed perturbation $\bar{h}_{\mu\nu}$) 
\begin{equation}
  f_{\nu}(h)=\partial^{\mu}\bar{h}_{\mu\nu}=
  \partial^{\mu}h_{\mu\nu}-\frac{1}{2}\partial_{\nu}h=
  0
\end{equation}
the Fadeev-Popov procedure lead us to add the gauge-fixing term $S_{GF}=-\frac{1}{2\xi}\int d^{4}x f_{\mu}^{2}$, so 
\begin{eqnarray}
  S&=&\int d^{4}x \frac{1}{2}h^{\mu\nu}{\cal O}_{\mu\nu,\rho\sigma}h^{\rho\sigma}
\end{eqnarray}
with 
\begin{eqnarray}
  {{\cal O}^{\mu\nu}}_{\rho\sigma}
  &=&({I^{\mu\nu}}_{\rho\sigma}-\eta^{\mu\nu}\eta_{\rho\sigma})\Box
  \nonumber\\
  &+&\eta^{\mu\nu}\partial_{\rho}\partial_{\sigma}
  +\eta_{\rho\sigma}\partial^{\mu}\partial^{\nu}
  -2\eta^{(\mu}_{(\rho}\partial_{}^{\nu)}\partial_{\sigma)}^{} \nonumber\\
  &+&\frac{1}{\xi}\Big(\frac{1}{4}\eta^{\mu\nu}\eta_{\rho\sigma}\Box
  -\frac{1}{2}\eta^{\mu\nu}\partial_{\rho}\partial_{\sigma}
  \nonumber\\
  &&\qquad-\frac{1}{2}\eta_{\rho\sigma}\partial^{\mu}\partial^{\nu}
  +\eta^{(\mu}_{(\rho}\partial_{}^{\nu)}\partial_{\sigma)}^{}\Big).\qquad
\end{eqnarray}
If we simplify by choosing the gauge parameter $1/\xi=2$, and define
the tensors
\begin{subequations}
  \begin{eqnarray}
    {I^{\mu\nu}}_{\rho\sigma}&\equiv&\frac{1}{2}(\delta^{\mu}_{\rho}\delta^{\nu}_{\sigma}+\delta^{\mu}_{\sigma}\delta^{\nu}_{\rho}), \\
    {J^{\mu\nu}}_{\rho\sigma}&\equiv&{I^{\mu\nu}}_{\rho\sigma}
    -\frac{1}{4}\eta^{\mu\nu}\eta_{\rho\sigma}, \\
    {K^{\mu\nu}}_{\rho\sigma}&=&
    \frac{1}{4}\eta^{\mu\nu}\eta_{\rho\sigma}
  \end{eqnarray}
\end{subequations}
(where ${I^{\mu\nu}}_{\rho\sigma}$, ${J^{\mu\nu}}_{\rho\sigma}$, and ${K^{\mu\nu}}_{\rho\sigma}$ are the identity operators on the space of symmetric matrices, symmetric traceless matrices, and pure-trace matrices, respectively, satisfying $I=J+K$), then can have 
\begin{subequations}
\begin{eqnarray}
  {{\cal O}^{\mu\nu}}_{\rho\sigma}
  &=&({I^{\mu\nu}}_{\rho\sigma}-\frac{1}{2}\eta^{\mu\nu}\eta_{\rho\sigma})\Box \\
  &=&{J^{\mu\nu}}_{\rho\sigma}\Box-{K^{\mu\nu}}_{\rho\sigma}\Box+i\epsilon {I^{\mu\nu}}_{\rho\sigma}\qquad
\end{eqnarray}
\end{subequations}
where in the second line we have made the substitution ${\cal O}_{\mu\nu\rho\sigma}\to{\cal O}_{\mu\nu\rho\sigma}+i\epsilon I_{\mu\nu\rho\sigma}$ to make the path integral $\int {\cal D}\varphi {\rm e}^{iS[h_{\mu\nu}]}$ converge.

Next we go to momentum space
\begin{eqnarray}
    {{\cal O}^{\mu\nu}}_{\rho\sigma}(\partial_{\mu}\to i k_{\mu})
    &=&{J^{\mu\nu}}_{\rho\sigma}(-k^{2}+i\epsilon) \nonumber\\
    &+&{K^{\mu\nu}}_{\rho\sigma}(+k^{2}+i\epsilon).\quad
\end{eqnarray}
The propagator is the operator ${{\cal D}^{\mu\nu}}_{\rho\sigma}$ satisfying
\begin{equation}
  \label{EH_propagator_definition}
  {{\cal O}^{\mu\nu}}_{\alpha\beta}{{\cal D}^{\alpha\beta}}_{\rho\sigma}=i {I^{\mu\nu}}_{\rho\sigma}.
\end{equation}
Since ${J^{\mu\nu}}_{\rho\sigma}$ and ${K^{\mu\nu}}_{\rho\sigma}$ are orthogonal projectors, it can be obtained by simply inverting their coefficients (and multiplying by $i$), so we obtain
\begin{subequations}
  \begin{eqnarray}
  \label{EH_propagator_prelim}
    {\cal D}_{\mu\nu\rho\sigma}^{{\rm EH}}&=&
    \frac{i J_{\mu\nu\rho\sigma}}{-k^{2}+i\epsilon}
    -\frac{i K_{\mu\nu\rho\sigma}}{-k^{2}-i\epsilon}
    .
  \end{eqnarray}
\end{subequations}
One can check that this matches the standard expression that can be found {\it e.g.} \cite{Donoghue:1995cz,Hinterbichler:2011tt} except that, again, since we have been careful to follow the precise $i\epsilon$'s that follow from the shift ${\cal O}_{\mu\nu\rho\sigma} \to {\cal O}_{\mu\nu\rho\sigma}+i\epsilon I_{\mu\nu\rho\sigma}$ we made to make the path integral converge, we find a $-i\epsilon$ in the denominator of the (gauge-dependent) term corresponding to the trace of $h_{\mu\nu}$ ($K_{\mu\nu\rho\sigma}$), whereas in the standard expression, this is replaced for simplicity by $+i\epsilon$, as in the denominator of the term corresponding to the trace-free part of $h_{\mu\nu}$ ($J_{\mu\nu\rho\sigma}$).

\subsubsection{Quadratic Gravity}
\label{quad_grav}

Now take the quadratic gravity (``QG'') action:
\begin{widetext}
\begin{subequations}
  \begin{eqnarray}
    S_{QG}&=&\!\int\!\! d^{4}x\sqrt{-g}\left[\frac{1}{16\pi G}
    R+\frac{1}{6f_{0}^{2}}R^{2}-\frac{1}{2f_{2}^{2}}C^{2}\right] \\
    &\approx&\!\int\!\! d^{4}x \Big\{\frac{1}{2\kappa^{2}}
    \Big[h^{\alpha\beta}\Box h_{\alpha\beta}
    -h\Box h
    +2h^{\alpha\beta}h_{,\alpha\beta}
    +2h^{\alpha}_{\gamma,\alpha}h_{\beta}^{\gamma,\beta}\Big]+\frac{({h^{\alpha\beta}}_{,\alpha\beta}-\Box h)^{2}}{6f_{0}^{2}}\nonumber\\
    &-&\frac{1}{2f_{2}^{2}}
    \Big[\frac{1}{2}(\Box h_{\alpha\beta})^{2}\!-\!\frac{1}{6}(\Box h)^{2}\!+\!\frac{1}{3}{h^{\alpha\beta}}_{,\alpha\beta}\Box h\!+\!\frac{1}{3}({h^{\alpha\beta}}_{,\alpha\beta})^{2}\!+\!h^{\alpha}_{\gamma,\alpha}\Box h_{\beta}^{\gamma,\beta}\Big]\Big\}
  \end{eqnarray}
\end{subequations}
\end{widetext}
where $\kappa=\sqrt{32\pi G}$ and, in this subsubsection, we take $h_{\mu\nu}$ to be dimensionless. The signs of the $R^2$ and $C^2$ terms are chosen to avoid tachyonic modes in the propagator.   Again, because the action is gauge-invariant under $h_{\mu\nu}\to h_{\mu\nu}+\partial_{\mu}\xi_{\nu}+\partial_{\nu}\xi_{\mu}$, we have to add a gauge-fixing term before we can invert the kinetic operator.  Instead of de Donder gauge ({\it i.e.} Lorentz gauge for the trace reversed perturbation $\bar{h}_{\mu\nu}$) 
it is now most convenient to apply Lorentz gauge to the {\it ordinary} metric perturbation $h_{\mu\nu}$:
\begin{equation}
  f_{\nu}(h)=\partial^{\mu}h_{\mu\nu}=0.
\end{equation}
And then, for quadratic gravity with its four derivatives, it is most natural to also add to the action a gauge fixing term which also has four derivatives: $S_{GF}=-\frac{1}{2\xi}\int d^{4}x f_{\mu}\Box f_{\mu}$, so
\begin{eqnarray}
  S&=&\int d^{4}x \frac{1}{2}h^{\mu\nu}{\cal O}_{\mu\nu,\rho\sigma}h^{\rho\sigma}
\end{eqnarray}
where ${{\cal O}^{\mu\nu}}_{\rho\sigma}$ is given by
\begin{equation}
    {{\cal O}^{\mu\nu}}_{\rho\sigma}
={{\cal O}_{{\rm EH}}^{\mu\nu}}_{\rho\sigma} +{{\cal O}_{(0)}^{\mu\nu}}_{\rho\sigma}
+{{\cal O}_{(2)}^{\mu\nu}}_{\rho\sigma}
+{{\cal O}_{(\xi)}^{\mu\nu}}_{\rho\sigma}
\end{equation} 
and ${{\cal O}_{EH}^{\mu\nu}}_{\rho\sigma}$, ${{\cal O}_{(0)}^{\mu\nu}}_{\rho\sigma}$, ${{\cal O}_{(2)}^{\mu\nu}}_{\rho\sigma}$,  
${{\cal O}_{(\xi)}^{\mu\nu}}_{\rho\sigma}$ are given by
\begin{widetext}
\begin{subequations}
\begin{eqnarray}
  {{\cal O}_{{\rm EH}}^{\mu\nu}}_{\rho\sigma}
  &\!\equiv\!&\frac{1}{\kappa^{2}}\Big[({I^{\mu\nu}}_{\rho\sigma}-\eta^{\mu\nu}\eta_{\rho\sigma})\Box+\eta^{\mu\nu}\partial_{\rho}\partial_{\sigma}
  +\eta_{\rho\sigma}\partial^{\mu}\partial^{\nu}
  -2{\eta^{(\mu}}_{(\rho}\partial^{\nu)}\partial_{\sigma)}\Big] \\
  {{\cal O}_{(0)}^{\mu\nu}}_{\rho\sigma}&\!\equiv\!&\frac{1}{3f_{0}^{2}} \Big[\partial^{\mu}\partial^{\nu}\partial_{\rho}\partial_{\sigma}+\eta^{\mu\nu}\eta_{\rho\sigma}\Box^{2}-\eta^{\mu\nu}\Box\partial_{\rho}\partial_{\sigma}-\eta_{\rho\sigma}\Box\partial^{\mu}\partial^{\nu}\Big]\\
  {{\cal O}_{(2)}^{\mu\nu}}_{\rho\sigma}&\!\equiv\!& 
  \frac{-1}{f_{2}^{2}}\Big[
  (\frac{1}{2}{I^{\mu\nu}}_{\rho\sigma}\!-\!\frac{1}{6}\eta^{\mu\nu}\eta_{\rho\sigma})\Box^{2}
  \!+\!\frac{1}{3}\partial^{\mu}\partial^{\nu}\partial_{\rho}\partial_{\sigma}
  \!+\!\frac{1}{6}(\eta^{\mu\nu}\Box\partial_{\rho}\partial_{\sigma}\!+\!\eta_{\rho\sigma}\Box\partial^{\mu}\partial^{\nu})
  \!-\!{\eta^{(\mu}}_{(\rho}\Box\partial^{\nu)}\partial_{\sigma)}\Big]\qquad\\
  {{\cal O}_{(\xi)}^{\mu\nu}}_{\rho\sigma}&\!\equiv\!&
  \frac{1}{\xi}{\eta^{(\mu}}_{(\rho}\Box
  \partial^{\nu)}\partial_{\sigma)}.
\end{eqnarray}
\end{subequations}
\end{widetext}
Now we go to momentum space ($\partial_{\mu}\to ik_{\mu}$), so that ${{\cal O}_{EH}^{\mu\nu}}_{\rho\sigma}$, ${{\cal O}_{(0)}^{\mu\nu}}_{\rho\sigma}$, ${{\cal O}_{(2)}^{\mu\nu}}_{\rho\sigma}$, and 
${{\cal O}_{(\xi)}^{\mu\nu}}_{\rho\sigma}$ become
\begin{subequations}
\begin{eqnarray}
  {{\cal O}_{EH}^{\mu\nu}}_{\rho\sigma}
  &=&\frac{-k^{2}}{\kappa^{2}}\Big[{{\cal P}_{(2)}^{\mu\nu}}_{\rho\sigma}
  -2{{\cal P}_{(0)}^{\mu\nu}}_{\rho\sigma}\Big] \qquad\\
  {{\cal O}_{(0)}^{\mu\nu}}_{\rho\sigma}&=&\frac{+k^{4}}{f_{0}^{2}}{{\cal P}_{(0)}^{\mu\nu}}_{\rho\sigma} \\
  {{\cal O}_{(2)}^{\mu\nu}}_{\rho\sigma}&=& 
  \frac{-k^{4}}{2f_{2}^{2}}{{\cal P}_{(2)}^{\mu\nu}}_{\rho\sigma} \\
  {{\cal O}_{(\xi)}^{\mu\nu}}_{\rho\sigma}&=&
  \frac{+k^{4}}{2\xi}\Big[{{\cal P}_{(1)\,\rho\sigma}^{\mu\nu}}+2{\tilde{{\cal P}}_{(0)\,\rho\sigma}^{\mu\nu}}\Big]
\end{eqnarray}
\end{subequations}
where we have defined the projection operators
\begin{subequations}
  \begin{eqnarray}
      {{\cal P}_{(2)}^{\mu\nu}}_{\rho\sigma}&=&
      2 {t^{(\mu}}_{(\rho}{t^{\nu)}}_{\sigma)}
      -\frac{1}{3}t^{\mu\nu}t_{\rho\sigma} \\
      {{\cal P}_{(1)}^{\mu\nu}}_{\rho\sigma}&=&
      2{t^{(\mu}}_{(\rho}{\ell^{\nu)}}_{\sigma)} \\
      {{\cal P}_{(0)}^{\mu\nu}}_{\rho\sigma}&=&\frac{1}{3}t^{\mu\nu}t_{\rho\sigma} \\
      {\tilde{\cal P}_{(0)\,\rho\sigma}^{\mu\nu}}&=&\ell^{\mu\nu}\ell_{\rho\sigma} 
  \end{eqnarray}
\end{subequations}
which are orthogonal and sum to the identity
$
    {I^{\mu\nu}}_{\rho\sigma}\!=\!{{\cal P}_{(2)}^{\mu\nu}}_{\rho\sigma}
    \!+\!{{\cal P}_{(1)}^{\mu\nu}}_{\rho\sigma}
    \!+\!{{\cal P}_{(0)\,\rho\sigma}^{\mu\nu}}\!+\!{\tilde {\cal P}_{(0)\,\rho\sigma}^{\mu\nu}}.
$
Putting this all together, we obtain
\begin{widetext}
\begin{subequations}
\begin{eqnarray}
    {\cal O}&=&
    -(\frac{k^{4}}{2f_{2}^{2}}+\frac{k^{2}}{\kappa^{2}}) {\cal P}_{}^{(2)}
    +(\frac{k^{4}}{f_{0}^{2}}+\frac{2k^{2}}{\kappa^{2}}){\cal P}_{}^{(0)}
    +\frac{k^{4}}{2\xi}({\cal P}_{}^{(1)}+2\tilde {\cal P}_{}^{(0)}) \\
    &=&-\frac{k^{2}(k^{2}+M_{2}^{2})}{2f_{2}^{2}}{\cal P}^{(2)}+\frac{k^{2}(k^{2}+M_{0}^{2})}{f_{0}^{2}}{\cal P}^{(0)}
    +\frac{k^{4}}{2\xi}({\cal P}^{(1)}+2\tilde {\cal P}^{(0)})\quad
\end{eqnarray}
\end{subequations}
\end{widetext}
where we have suppressed $\mu\nu\rho\sigma$ indices for brevity, and defined the mass parameters $M_{0}$ and $M_{2}$:
\begin{subequations}
  \begin{eqnarray}
    M_{0}^{2}&\equiv &2f_{0}^{2}/\kappa^{2}=f_{0}^{2}/(16\pi G) \\
    M_{2}^{2}&\equiv &2f_{2}^{2}/\kappa^{2}=f_{2}^{2}/(16\pi G).
  \end{eqnarray}
\end{subequations}

Now we make the substitution ${\cal O}_{\mu\nu\rho\sigma}\to{\cal O}_{\mu\nu\rho\sigma}+i\epsilon I_{\mu\nu\rho\sigma}$ to make the path integral $\int {\cal D}\varphi {\rm e}^{iS_{QG}[h_{\mu\nu}]}$ converge, and generalize $i\epsilon I$ as
\begin{equation}
  i\epsilon I\to i\epsilon_{0}{\cal P}^{(0)}+i\tilde\epsilon_{0}\tilde {\cal P}^{(0)}+i\epsilon_{1}{\cal P}^{(1)}+i\epsilon_{2}{\cal P}^{(2)},
\end{equation}
with $\epsilon_{0},$  $\tilde{\epsilon_{0}}$, $\epsilon_{1}$ and $\epsilon_2$ infinitesimal and positive. 

Since the various $\epsilon$ factors are formal parameters whose magnitude is irrelevant, we will henceforth allow ourselves to rescale them arbitrarily while preserving their sign. In subsequent expressions we will cease to distinguish them and just refer to them all as ``$\epsilon$."

Then, to first order in the $\epsilon$, we have
\begin{eqnarray}
  {\cal O}&=&-\frac{(k^{2}-i\epsilon\kappa^{2})(k^{2}+M_{2}^{2}+i\epsilon\kappa^{2})}{2f_{2}^{2}}{\cal P}^{(2)}\nonumber\\
  &&+\frac{(k^{2}+i\epsilon\kappa^{2})(k^{2}+M_{0}^{2}-i\epsilon \kappa^{2})}{f_{0}^{2}}{\cal P}^{(0)}\nonumber\\
  &&+\frac{(k^{2}+i\epsilon\xi)^{2}}{2\xi}[{\cal P}^{(1)}+2\tilde {\cal P}^{(0)}].
\end{eqnarray}

Then the propagator ${{\cal D}^{\mu\nu}}_{\rho\sigma}$ satisfying
\begin{equation}
  {{\cal O}^{\mu\nu}}_{\alpha\beta}{{\cal D}^{\alpha\beta}}_{\rho\sigma}=i {I^{\mu\nu}}_{\rho\sigma}.
\end{equation}
is again given by inverting the projector coefficients and multiplying by $i$: 
\begin{eqnarray}
  \label{QG_propagator_prelim}
    {\cal D}_{QG}&=&
    -\frac{2if_{2}^{2}}
    {(k^{2}-i\epsilon\kappa^{2})(k^{2}+M_{2}^{2}+i\epsilon\kappa^{2})}{\cal P}^{(2)}
    \nonumber \\
    &&+\frac{i f_{0}^{2}}{(k^{2}+i\epsilon\kappa^{2})(k^{2}+M_{0}^{2}-i\epsilon\kappa^{2})}{\cal P}^{(0)}
    \nonumber\\
    &&+\frac{2i\xi}{(k^{2}+i\epsilon\xi)^{2}}[{\cal P}^{(1)}+\frac{1}{2}\tilde {\cal P}^{(0)}]
\end{eqnarray}
Choosing the gauge parameter $\xi=0$, and decomposing in partial fractions, this becomes
\begin{eqnarray}
  \label{QG_propagator_prelim}
    {\cal D}_{QG}&\!=\!&
    \kappa^{2}\Big\{
    \frac{i{\cal P}^{(2)}}{-k^{2}+i\epsilon\kappa^{2}}
    \!-\!\frac{\frac{1}{2}i{\cal P}^{(0)}}{-k^{2}-i\epsilon\kappa^{2}}
     \nonumber\\
    &\!-\!&\frac{i {\cal P}^{(2)}}{-k^{2}-M_{2}^{2}-i\epsilon\kappa^{2}}
    \!+\!\frac{\frac{1}{2}i {\cal P}^{(0)}}{-k^{2}-M_{0}^{2}+i\epsilon\kappa^{2}}\Big\}.\qquad\quad
\end{eqnarray}
Let us summarize the content of this expression.  The two terms on the 
first line of (\ref{QG_propagator_prelim}) both have their poles at $k^{2}=0$, and combine to reproduce the standard Einstein-Hilbert propagator: in particular, the first term ($\propto {\cal P}^{(2)}$) is gauge-independent and describes the usual spin-2 massless time-ordered Einstein-Hilbert graviton, while the second term ($\propto{\cal P}^{(0)}$) is a gauge-dependent term.  On the second line of (\ref{QG_propagator_prelim}): the first term ($\propto {\cal P}^{(2)}$) describes a spin-2 massive {\it anti}-time-ordered ghost graviton with mass squared $M_{2}^{2}=f_{2}^{2}M_{pl}^{2}/2$ (it is a ghost because of the overall minus sign in front, and anti-time-ordered because of the $-i\epsilon\kappa^{2}$ in the denominator); and the second term ($\propto {\cal P}^{(0)}$) describes a spin-0 massive time-ordered graviton with mass squared $M_{0}^{2}=f_{0}^{2}M_{pl}^{2}/2$.  If $f_{2}^{2}<0$, the massive spin-2 graviton becomes a tachyon, and if $f_{0}^{2}<0$, the massive spin-0 graviton becomes a tachyon.


\subsubsection{Pure Quadratic Gravity}

Finally, consider the special case of ``pure quadratic gravity'' (``PQG'') -- {\it i.e.} the special case of quadratic gravity in which $\kappa\to\infty$ (so the Einstein-Hilbert term disappears from the action, and we only have the quadratic curvature terms, with their dimensionless coefficient $f_{0}^{-2}$ and $f_{2}^{-2}$.\footnote{In the pure-quadratic-gravity limit, another interesting possibility is to consider the opposite sign for the $R^2$ coupling $f_{0}^{2}$.  Taking this sign for $f_{0}^{2}$, and the "conformally-flat limit" $f_{2}^{2}\to 0$ for the $C^{2}$ coupling, the resulting theory is an example of a positive and asymptotically free higher-derivative theory \cite{Bateman:2026eyj, Anderson:2026ilf}.}  Then we have, to first order in the $\epsilon$'s,
\begin{eqnarray}
  {\cal O}&=&-\frac{(k^{2}-i\epsilon f_{2}^{2})^{2}}{2f_{2}^{2}}{\cal P}^{(2)}
  +\frac{(k^{2}+i\epsilon f_{0}^{2})^{2}}{f_{0}^{2}}{\cal P}^{(0)} 
  \nonumber\\
  &&+\frac{(k^{2}+i\epsilon\xi)^{2}}{2\xi}[{\cal P}^{(1)}+2\tilde {\cal P}^{(0)}]\quad
\end{eqnarray}
where the $\epsilon$'s are defined as above (in the Appendix \ref{quad_grav}).  Then the propagator satisfying
\begin{equation}
  {{\cal O}^{\mu\nu}}_{\alpha\beta}{{\cal D}^{\alpha\beta}}_{\rho\sigma}=i {I^{\mu\nu}}_{\rho\sigma}
\end{equation}
is given by inverting the coefficients of each projector and multiplying by $i$: 
\begin{equation}
  {\cal D}=-\frac{2if_{2}^{2}{\cal P}^{(2)}}{(-k^{2}+i\epsilon f_{2}^{2})^{2}}
  +\frac{if_{0}^{2}{\cal P}^{(0)}}{(-k^{2}-i\epsilon f_{0}^{2})^{2}}
\end{equation}
where we have again chosen the gauge $\xi=0$.

\section{Fermion-antifermion annihilation cross section in quadratic gravity}
\label{Fermion_Graviton_Vertex}

In this Appendix, we compute the cross section in quadratic gravity for a Standard Model fermion-antifermion pair to gravitationally annihilate into a pair of dark matter particles (see Fig.~\ref{feynman diagram}).

We use the particle physics convention of mostly minus metric in appendices B, C, and D. The stress-tensor for the Dirac action $S = \int d^4x\sqrt{-g}\,\overline{\psi}\,(i\slashed{D}-m)\psi$ gives the fermion-antifermion-graviton interaction vertex \cite{Pastor-Gutierrez:2024sbt}:
\begin{eqnarray}
    \Gamma^{\alpha\beta}(p_{\psi}, q_{\bar{\psi}}) = \frac{-i}{8}\biggr[4m\eta^{\alpha\beta} + \gamma^{\alpha}(p^{\beta}-q^{\beta}) + \nonumber\\\gamma^{\beta}(p^{\alpha} - q^{\alpha}) - 2\eta^{\alpha\beta}(\slashed{p}-\slashed{q})\biggr]\quad
\end{eqnarray}
where $p$ and $q$ are the incoming four momentum of the fermion and anti-fermion pair. We first calculate the amplitude coming from the scalar mode of the propagator:
\begin{eqnarray}
    i\mathcal{M}^{(0)}&=&\overline{v}(q)\Gamma^{\mu\nu}(p,q)u(p)\frac{f_0^2i}{(p+q)^4}\mathcal{P}^{(0)}_{\mu\nu\alpha\beta}\nonumber\\
    &&\qquad\times \overline{u}(p')\Gamma^{\alpha\beta}(p',q')v(q')\,.
\end{eqnarray}
Let us simplify this using $\mathcal{P}^0_{\mu\nu\alpha\beta} = \frac{1}{3}t_{\mu\nu}t_{\alpha\beta}$; the contraction of the tensor $t_{\alpha\beta}$ coming from the propagator with the dark matter-graviton vertex is given by:
\begin{eqnarray}
    &&t_{\alpha\beta}\Gamma^{\alpha\beta}(p',q') = \frac{-i}{8}\biggr(\eta_{\alpha\beta} - \frac{(p'+q')_{\alpha}(p'+q')_{\beta}}{(p'+q')^2}\biggr)\nonumber\\
    &&\qquad\times\big(4 m_{dm}\eta^{\alpha\beta}\!+\! 2\gamma^{(\alpha}(p'\!-\!q')^{\beta)}\!-\! 2\eta^{\alpha\beta}(\slashed{p}'\!-\! \slashed{q}') \big)\nonumber\\
    &&\Rightarrow\overline{u}(p')t_{\alpha\beta}\Gamma^{\alpha\beta}v(q')=  \frac{-i}{2}m_{dm}\overline{u}(p')v(q')
\end{eqnarray}
where $m_{dm}$ is the mass of the dark matter, and we have used the equation of motion $(\slashed{q}' + m_{dm})v(q') = 0$ for the dark matter particle to obtain the last line. An analysis similar to the one for the fermion-graviton vertex gives the following value for the amplitude coming from the scalar mode of the graviton propagator:
\begin{eqnarray}
    i\mathcal{M}^{(0)} = \frac{if_0^2}{3(p+q)^4}\left(-\frac{i}{2}m_{\psi}\bar v(q)u(p)\right)\nonumber\\
    \times \left(-\frac{i}{2}m_{dm}\bar u(p')v(q')\right)
\end{eqnarray}
Note that $\frac{m_{dm}m_{\psi}}{s} << 1$ (where $s = (p+q)^2$) and therefore the scalar amplitude will be negligible compared to the tensor amplitude given by:
\begin{eqnarray}
    i\mathcal{M}^{(2)}&=& \overline{v}(q)\Gamma^{\mu\nu}(p,q)u(p)\frac{-2f_2^2 i}{(p+q)^4}\mathcal{P}^{(2)}_{\mu\nu\alpha\beta}\nonumber\\
    &&\qquad \times \overline{u}(p')\Gamma^{\alpha\beta}(p',q')v(q')\,.
\end{eqnarray}
We use $\mathcal{P}^{(2)}_{\mu\nu\alpha\beta} = \frac{1}{2}(t_{\mu\alpha}t_{\nu\beta}+t_{\mu\beta}t_{\nu\alpha}) - \frac{1}{3}t_{\mu\nu}t_{\alpha\beta} = \frac{1}{2}(t_{\mu\alpha}t_{\nu\beta}+t_{\mu\beta}t_{\nu\alpha}) - \mathcal{P}_{\mu\nu\alpha\beta}^{(0)}$ to simplify the expression above. We have already done the calculation with the spin$-0$ projector. The remaining part simplifies to the following upon using the equation of motion for the external states:
\begin{eqnarray}
    &&\bar v(q)\Gamma^{\mu\nu}u(p)\frac{(t_{\mu\beta}t_{\nu\alpha} + t_{\mu\alpha}t_{\nu\beta})}{2}\bar u(p')\Gamma^{\alpha\beta}v(q') = \nonumber\\
    &&\biggr(\frac{-i}{8}\biggr)^2\bar v(q)\biggr[\gamma_{\alpha}(p-q)_{\nu} + \gamma_{\nu}(p-q)_{\alpha}\biggr]u(p)\nonumber\\
    &&\times\bar u(p')\biggr[\gamma^{\alpha}(p'-q')^{\nu} + \gamma^{\nu}(p'-q')^{\alpha}\biggr]v(q').
\end{eqnarray}
The total amplitude is given by $i\mathcal{M} = i\mathcal{M}^{(0)} + i\mathcal{M}^{(2)}$. Since $\frac{m_{dm}m_{\psi}}{s} <<1$, the spin$-0$ contribution is negligible. To calculate $|\mathcal{M}|^2$, we average the spin of the initial particles and sum over the spin of the final particles using the identity $\sum\limits_{s}u(p,s)\bar u(p,s) = \slashed{p} + m_{\psi}$, $\sum\limits_{s}v(q,s)\bar v(q,s) = \slashed{q} - m_{\psi}$. Ignoring the particle's mass compared to their momentum, we get:
\begin{eqnarray}
    &|\mathcal{M}|^2 = \frac{f_2^4}{(64s^2)^2}\text{tr}\big[\slashed{q}(\gamma_{\nu}(p-q)_{\alpha} + \gamma_{\alpha}(p-q)_{\nu})\slashed{p}\qquad\nonumber\\
    &(\gamma_{\mu}(p-q)_{\beta} + \gamma_{\beta}(p-q)_{\mu})\big]\text{tr}\big[\slashed{q}'(\gamma^{\nu}(p'-q')^{\alpha} +\quad \nonumber\\
    &\gamma^{\alpha}(p'-q')^{\nu})\slashed{p}'(\gamma^{\mu}(p'-q')^{\beta} + \gamma^{\beta}(p'-q')^{\mu})\big]\qquad
\end{eqnarray}
where it is useful to introduce:
\begin{eqnarray}
    &s = (p+q)^2 \approx 2p\cdot q = 2p'\cdot q'\nonumber\\
    &t = (p+p')^2 \approx 2p\cdot p' = 2q\cdot q' \nonumber\\
    &u = (p+q')^2 \approx 2p\cdot q' = 2q\cdot p' \nonumber\\
    &s+t+u \approx 0
\end{eqnarray}
Using the trace identity for Clifford matrices leads to further simplifications. We get:
\begin{eqnarray}
    |\mathcal{M}|^2\!&=&\!\frac{f_2^4}{256s^4}\bigg[2(t\!-\!u)^2(u^2\!+\!t^2)\!+\! (s^2\!-\!(t\!-\!u)^2)^2 \nonumber\\ 
    &&\qquad\quad+ 2(t-u)^2((t-u)^2-s^2) \bigg]
\end{eqnarray}
We can then eliminate $u$ in terms of $s$ and $t$ and rewrite the expression in terms of the scattering angle $\theta$ for 2-to-2 scattering in the massless limit (using $cos\,\theta=\frac{2t}{s} + 1$) to get:
\begin{equation}
  |\mathcal{M}|^{2}=\frac{f_{2}^{4}}{256}
  (4\,{\rm cos}^{4}\theta-3\,{\rm cos}^{2}\theta+1).
\end{equation}

Using this in the expression for the differential cross-section for 2-to-2 massless scattering \cite{Schwartz:2014sze}:
\begin{eqnarray}
  \label{genera_2to2}
    \frac{d\sigma}{d\Omega} = \frac{|\mathcal{M}|^2}{64\pi^2s},
\end{eqnarray}
we can write $d\Omega = 2\pi d(cos(\theta))$ and perform the angular integral to obtain the total cross section
\begin{eqnarray}
    \sigma =
    \frac{f_{2}^{4}}{5120\pi s}
    =\frac{\frac{3}{4}f_2^4}{3840\pi s}.
\end{eqnarray}
for annihilating spin $1/2$ fermions.

\section{Scalar-scalar annihilation cross section in quadratic gravity}
\label{Scalar_Graviton_Vertex}

In this Appendix, we compute the cross section in quadratic gravity for a scalar-scalar pair to gravitationally annihilate into a pair of dark matter particles (see Fig.~\ref{feynman diagram}).

We have seen already that the dark-matter graviton vertex will be proportional to the mass of the dark matter particle for the scalar mode and will not contribute. We further find that the vertex coupling a conformally coupled scalar field with the scalar mode of the graviton is zero. Therefore only the tensor mode contributes to the amplitude\footnote{This is because the variation of the matter action \textit{w.r.t.} the scalar mode of graviton gives the trace of the stress-energy tensor}. Consider a scalar field with non-minimal coupling $\xi$, such that $S = \int d^4x\sqrt{-g}(-\frac{1}{2}\phi(\Box + \xi R)\phi)$. The vertex rule for two scalars and a graviton comes from the scalar stress tensor and is given by:
\begin{eqnarray}
    &&\Gamma^{\alpha\beta}_0 = i\bigg[\frac{1}{2}(p^{\alpha}q^{\beta} + p^{\beta}q^{\alpha} - \eta^{\alpha\beta}p\cdot q) + \xi(\eta^{\alpha\beta}(2p\cdot q \nonumber\\&&+ p^2 + q^2)-p^{\alpha}q^{\beta}-q^{\alpha}p^{\beta}-p^{\alpha}p^{\beta}-q^{\alpha}q^{\beta})\bigg]\,.
\end{eqnarray}
The amplitude for the scalar mode is given by:
\begin{eqnarray}
    i\mathcal{M}^{(0)} = \Gamma_0^{\mu\nu}(p,q)\frac{f_0^2 i}{(p+q)^4}\mathcal{P}^{(0)}_{\mu\nu\alpha\beta}\bar u(p')\nonumber\\\Gamma^{\alpha\beta}(p',q')v(q').
\end{eqnarray}
The polarization tensor for the scalar mode is given by $\mathcal{P}^{0}_{\mu\nu\alpha\beta} = \frac{1}{3}t_{\mu\nu}t_{\alpha\beta}$. The scalar-graviton vertex is proportional to $\Gamma_0^{\mu\nu}t_{\mu\nu} = (6\xi - 1)p\cdot q$. Therefore, it vanishes for conformal coupling.

Let us turn now to the amplitude for the tensor mode, given by:
\begin{eqnarray}
    i\mathcal{M}^{(2)} = \Gamma^{\mu\nu}_0(p,q)\frac{-2f_2^2 i}{(p+q)^4}\mathcal{P}^{(2)}_{\mu\nu\alpha\beta}\overline{u}(p')\nonumber\\\times \Gamma^{\alpha\beta}(p',q')v(q')\,.
\end{eqnarray}
Again, we use $\mathcal{P}^{(2)}_{\mu\nu\alpha\beta} = \frac{1}{2}(t_{\mu\alpha}t_{\nu\beta}+t_{\mu\beta}t_{\nu\alpha}) - \mathcal{P}_{\mu\nu\alpha\beta}^{(0)}$ and ignore the scalar projector. The amplitude squared is given by:
\begin{eqnarray}
    &&|\mathcal{M}|^2\!=\! \frac{4f_2^4}{64s^4}\Gamma_0^{\mu\nu}\Gamma_0^{\alpha\beta}\text{tr}\big[\slashed{q}'(\gamma_{\alpha}(p'\!-\!q')_{\beta}\!+\!\gamma_{\beta}(p'\!-\!q')_{\alpha})\nonumber\\  &&\qquad\qquad\quad\times\slashed{p}'(\gamma_{\mu}(p'-q')_{\nu} + \gamma_{\nu}(p'-q')_{\mu})\big].
\end{eqnarray}
For the ease of calculation, we work with the minimal coupling case $\xi = 0$, note that the amplitude coming from the scalar mode will remain negligible due to the relation $m^2_{dm}/s << 1$. Upon using the trace identity for the gamma matrices and expanding we get:
\begin{eqnarray}
    |\mathcal{M}|^2 = \frac{f_2^4}{32s^4}(2t+s)^2(s^2-(2t+s)^2).
\end{eqnarray}
This can again be written in terms of the scattering angle using $cos\,\theta=\frac{2t}{s} + 1$ to get:
\begin{eqnarray}
    |\mathcal{M}|^2 = \frac{f_2^4}{32}cos^2(\theta)(1-cos^2\theta).
\end{eqnarray}
Substituting this into Eq.~(\ref{genera_2to2}) and integrating over $d\Omega$ now yields the following expression for the total cross section:
\begin{eqnarray}
    \sigma = \frac{f_2^4}{3840\pi s}
\end{eqnarray}
for annihilating spin $0$ scalars.

\section{Gauge boson annihilation cross section in quadratic gravity}
\label{Photon_Graviton_Vertex}

In this Appendix, we compute the cross section in quadratic gravity for a pair of Standard Model gauge bosons to gravitationally annihilate into a pair of dark matter particles (see Fig.~\ref{feynman diagram}).

To the order we work here, we shall only need the quadratic terms in the action for all the gauge bosons of the Standard Model.  

The vertex for two gauge bosons and a graviton is given by the gauge boson stress-tensor:
\begin{eqnarray}
    &&{\Gamma^{\alpha\beta}}_{\gamma\delta}(p,q)\!=\!i\Big[\frac{1}{2}\eta^{\alpha\beta}(\eta_{\gamma\delta}p\cdot q\!-\!p_{(\gamma}q_{\delta)})\!-\!p^{(\alpha}q^{\beta)}\eta_{\gamma\delta} \nonumber\\
    &&-(p\cdot q){\eta^{(\alpha}}_{\!(\gamma}{\eta^{\beta)}}_{\!\delta)}\!+\!{\eta^{(\alpha}}_{(\gamma}p^{\beta)}q_{\delta)}\!+\!{\eta^{(\alpha}}_{\!(\gamma}q^{\beta)}p_{\delta)}\Big].
\end{eqnarray}
Again, we can ignore the amplitude coming from the scalar mode of the graviton because the dark-matter vertex is proportional to the dark matter particle's mass. The amplitude is thus given by:
\begin{eqnarray}
    i\mathcal{M}^{(2)} = \xi_{\alpha}(p)\tilde\xi_{\beta}(q)\Gamma^{\alpha\beta\gamma\delta}(p,q)\frac{-2if_2^2}{(p+q)^4}\nonumber\\\times\mathcal{P}^{(2)}_{\gamma\delta\mu\nu}\bar u(p')\Gamma^{\mu\nu}(p',q')v(q')
\end{eqnarray}
where $\xi_{\alpha}(p)$ and $\tilde\xi_{\beta}(q)$ are the polarization vectors for external photons. For on-shell massless vector bosons we get $\xi\cdot p = 0$ and $\tilde\xi\cdot q = 0$, this leads to many simplifications.  We find:

\begin{eqnarray}
    &&\xi_{\alpha}(p)\tilde\xi_{\beta}(q)\Gamma^{\alpha\beta}_{\gamma\delta}(p,q) = \frac{i}{2}\bigg[\xi\cdot\tilde\xi(\eta_{\gamma\delta}p\cdot q-p_{(\gamma}q_{\delta)})\nonumber\\
    &&\qquad\qquad\qquad-(\xi\!\cdot\! q)(\tilde\xi\cdot p)\eta_{\delta\gamma}\!-\! 2p\!\cdot\! q(\xi_{(\gamma}\tilde\xi_{\delta)})\nonumber\\
    &&\qquad\qquad\qquad+(\xi\!\cdot\! q)\tilde\xi_{(\delta}p_{\gamma)}\!+\! (\tilde\xi\!\cdot\! p)\xi_{(\gamma}q_{\delta)}\bigg].\qquad
\end{eqnarray}
To calculate the square of the amplitude, we will have to take the average over the photon polarization using the identity $\sum\limits_{\lambda}\xi^{(\lambda)}_{\mu}(p)\xi^{(\lambda)}_{\nu}(p) = -\eta_{\mu\nu}$, $\sum\limits_{\lambda}\tilde\xi^{(\lambda)}_{\mu}(q)\tilde\xi^{(\lambda)}_{\nu}(q) = -\eta_{\mu\nu}$ in Feynman gauge.  After this simplification, $|\mathcal{M}|^{2}$ is given by:
\begin{eqnarray}
    |\mathcal{M}|^2 
    &=& \frac{f_2^4}{64s^4}\bigg[\frac{3}{4}(2t+s)^2(s^{2}-(2t+s)^2) + 2s^4 \nonumber\\&&\qquad\qquad+ \frac{1}{4}(2t(t+s)- (2t+s)^2)s^{2}\bigg] \nonumber\\ 
    &=& \frac{f_2^4}{512}\bigg[15+5\,{\rm cos}^{2}\theta-6\,{\rm cos}^{4}\theta\bigg]
\end{eqnarray}
where in the final expression we have eliminated $u$ and used $cos\,\theta=\frac{2t}{s} + 1$.  Substituting the result into Eq.~(\ref{genera_2to2}) and integrating over $d\Omega$ yields the following expression for the total cross section:
\begin{eqnarray}
    \sigma =\frac{29 f_{2}^{4}}{15360\pi s}
    =\frac{\frac{29}{4}f_2^4}{3840\pi s}
\end{eqnarray}
for annihilating spin 1 gauge bosons.

\section{Gravitational annihilation of thermal bath particles into dark matter}
\label{App:Gamma_Bar}

In this Appendix, we combine the results of Appendices \ref{Fermion_Graviton_Vertex}, \ref{Scalar_Graviton_Vertex}, and \ref{Photon_Graviton_Vertex} to obtain the average rate $\overline{\Gamma}$ of 2-to-2 annihilation of thermal bath particles into dark matter.  The answers simplify both in the UV (where the graviton propagator reduces to that of pure quadratic gravity, $\propto k^{-4}$), and in the IR (where the graviton propagator reduces to that of Einstein gravity, $\propto k^{-2}$).

\subsection{General formalism}

The evolution of the dark matter number density $n_{dm}$, due to the 2-to-2 gravitational annihilation process $i+\bar{i}\to DM+DM$ shown in Fig.~\ref{feynman diagram} is governed by the equation \cite{Kolb:1990vq}
\begin{equation}
  \label{ndot_dm_v1}
  \dot{n}_{dm}+3Hn_{dm}=2\sum_{i}n_{i}^{2}\langle \sigma v\rangle_{i}
\end{equation}
where there is a factor of $2$ on the right-hand side because each annihilation yields two dark matter particles in the final state; the sum runs over all species $i$ in equilibrium with the thermal bath (including both different particle types and different spin states); $\langle\sigma v\rangle_{i}$ is the thermal average at temperature $T$ of the annihilation cross section $\sigma$ times velocity $v$ for species $i$; $n_{i}$ is the equilibrium number density of the species $i$, which is given by \cite{Kolb:1990vq}
\begin{equation}
  \label{n_i_thermal}
    n_{i}=\xi_{i}\frac{\zeta(3)}{\pi^{2}}T^{3}
\end{equation}
where $\xi_{i}$ is $1$ or $3/4$ if the species $i$ is bosonic or fermionic, respectively; and -- assuming the chemical potential is negligible at high temperatures -- this is the same as the number density of the corresponding anti-species $\bar{i}$ \cite{Kolb:1990vq}. 

Multiplying both sides by $a^3$, Eq.~(\ref{ndot_dm_v1}) becomes
\begin{equation}
  \label{ndot_dm_v2}
  (a^{3}n_{dm})\,\dot{}=2a^{3}\sum_{i}n_{i}^{2}\langle\sigma v\rangle_{i}.
\end{equation}
Integrating with respect to time, this becomes
\begin{eqnarray}
  \label{ndot_dm_v3}
    (a^{3}n_{dm})_{f}
    &=&2\int dt\sum_{i}a^{3}n_{i}^{2}\langle \sigma v\rangle_{i} \nonumber\\ 
    &=&2\int dt\sum_{i}(a^{3}n_{i})\Gamma_{i}
\end{eqnarray}
where $(a^{3}n_{dm})_{f}$ denotes the final value of $a^{3}n_{dm}$, right after the era of dark matter production has ended (and we have assumed that the initial value of $a^{3}n_{dm}$ was zero); and 
\begin{equation}
  \label{def_Gamma_i}
  \Gamma_{i}=n_{i}\langle\sigma v\rangle_{i}
\end{equation} 
is the annihilation rate for species $i$. Note that the factor $a^{3}n_{i}$ is approximately constant over the relatively brief period when the dark matter is dominantly produced (assuming this took place deep in the radiation era, when the temperature $T$ was well above the masses of the all or nearly all of the particle species in the thermal bath).

If we divide both sides of (\ref{ndot_dm_v3}) by 
\begin{equation}
    a^{3}n_{tot}=a^{3}\sum_{i}n_{i}
\end{equation}
where $n_{tot}$ is the total number density of particles in the thermal bath (and $a^{3}n_{tot}$ is again approximately constant over the brief period when the dark matter is dominantly produced), we obtain an expression for the number of dark matter particle produced, per particle in the thermal bath, just after the era of dark matter production has ended:
\begin{equation}
  \label{n_dm_over_n_tot}
    \frac{n_{dm}}{n_{tot}}=\frac{(a^{3}n_{dm})_{f}}{a^{3}n_{tot}}
    =2\int dt \overline{\Gamma}
\end{equation}
where
\begin{equation}
  \label{def_Gamma_bar}
  \overline{\Gamma}=\frac{\sum_{i}a^{3}n_{i}\Gamma_{i}}{\sum_{i}a^{3}n_{i}}  
\end{equation}
is the average annihilation rate, averaged over all species in the thermal bath.  Using (\ref{n_i_thermal}), this becomes
\begin{equation}
  \label{expr_Gamma_bar} 
  \overline{\Gamma}=\frac{g_{\ast n}^{(0)}\Gamma_{0}+
  g_{\ast n}^{(\frac{1}{2})}\Gamma_{\frac{1}{2}}+g_{\ast n}^{(1)}\Gamma_{1}}{g_{\ast n}}
\end{equation}
where $g_{\ast n}$ denotes the effective number of relativistic species in the thermal bath:
\begin{equation}
    g_{\ast n}=\sum_{i}\xi_{i}
\end{equation}
and $g_{\ast n}^{(a)}$ denotes the contribution to $g_{\ast n}$ from particles with spin $a$:
\begin{equation}
    g_{\ast n}^{(a)}=\sum_{i={\rm spin}\,a}\xi_{i}.
\end{equation}
For the Standard Model including a right-handed neutrino in each generation ({\it i.e.}, three right-handed neutrinos in total, but only two of which thermalize  while the third is dark matter and only interacts gravitationally) we have
\begin{subequations}
  \label{gstar_n_sm}
    \begin{eqnarray}
        g_{\ast n}^{(0)}&=&4 \\
        g_{\ast n}^{(\!\frac{1}{2})}&=&\frac{3}{4}
        (2\times 16+1\times 15)\times 2=\frac{141}{2}\qquad \\
        g_{\ast n}^{(1)}&=&2\times 12=24 \\
        g_{\ast n}&=&\frac{197}{2}\;.
    \end{eqnarray}
\end{subequations}

\subsection{Computing $\langle \sigma v\rangle_{i}$ and $\Gamma_{i}$}

In this Appendix, we combine the results of Appendices \ref{Fermion_Graviton_Vertex}, \ref{Scalar_Graviton_Vertex}, and \ref{Photon_Graviton_Vertex} to 
compute the thermally averaged quantities $\langle\sigma v\rangle_{i}$ and $\Gamma_{i}$ appearing in Eq.~(\ref{def_Gamma_i}).  We can approximate $\langle\sigma v\rangle_{i}$ by the formula given in \cite{Gondolo:1990dk}:
\begin{eqnarray}
  \label{sigma v}
    \langle\sigma v\rangle_{i} &=& \frac{1}{8m^4TK_{2}^2(m/T)}\int\limits_{4m^2}^{\infty}\sigma_{i}(s)(s-4m^2)\sqrt{s}\nonumber\\
    &&\qquad\qquad\qquad\qquad \times K_1(\sqrt{s}/T)ds
\end{eqnarray}
where, $K_\nu(x)$ is the modified Bessel function of the second kind of order $\nu$.  (An even more precise answer could be obtained by evaluating $\langle\sigma v\rangle_{i}$ numerically -- we leave this to future work.) 

We want the annihilation rate in the very early universe, when the masses of the particles are negligible relative to the temperature $T$.  So we can simplify (\ref{sigma v}) by taking the $m\to0$ limit. 
First let us consider the prefactor $\frac{1}{8m^4TK_{2}^2(m/T)}$ in (\ref{sigma v}): using the Laurent expansion of $K_2(z)$ around $z = 0$
\begin{eqnarray}
    K_2(z) = \frac{2}{z^2} -\frac{1}{2} + \mathcal{O}(z^2)
\end{eqnarray}
we have
\begin{eqnarray}
  \label{prefactor}
    \lim\limits_{m\rightarrow 0} \frac{1}{8m^4TK_{2}^2(m/T)} = \frac{1}{32T^5}.
\end{eqnarray}

Next consider the integral in (\ref{sigma v}).  This depends on whether we use the expression for $\sigma(s)$ obtained: (i) in the UV (where the graviton propagator reduces to that of pure quadratic gravity, $\propto k^{-4}$), or (ii) in the IR (where the graviton propagator reduces to the one from Einstein gravity, $\propto k^{-2}$).  We consider these two regimes separately.

\subsubsection{UV calculation}

We have seen in the previous appendices that, in the UV limit (described by pure quadratic gravity), the gravitational annihilation cross section of thermal bath particles into dark matter becomes
\begin{equation}
    \sigma_{i}(s) = \frac{A_{i}^{\rm UV}f_2^4}{s}\
\end{equation} 
where the constant $A_{i}^{\rm UV}$ depends on the spin of the annihilating species $i$. In particular, in Appendices \ref{Fermion_Graviton_Vertex}, \ref{Scalar_Graviton_Vertex}, and \ref{Photon_Graviton_Vertex}, we found
\begin{subequations}
  \begin{eqnarray}
    A_i^{\rm UV}&=&\frac{1}{3840\pi}\qquad(i={\rm spin}\;0) \\
    A_{i}^{\rm UV}&=&\frac{3/4}{3840\pi}\qquad(i={\rm spin}\,\frac{1}{2}) \\
    A_{i}^{\rm UV}&=&\frac{29/4}{3840\pi}\qquad(i={\rm spin}\;1).
  \end{eqnarray}
\end{subequations}
If we define $x = \frac{\sqrt{s}}{T}$ then in the limit $m\rightarrow 0$, the integral in (\ref{sigma v}) becomes
\begin{eqnarray}
    &&\lim_{m\rightarrow 0}\int\limits_{4m^2}^{\infty}\sigma_{i}(s)(s-4m^2)\sqrt{s}\, K_1(\sqrt{s}/T)ds \nonumber\\ 
    &&=2A_i^{\rm UV}f_2^4T^3\int\limits_0^{\infty}x^2K_1(x)dx = 4A_i^{\rm UV}f_2^4T^3.\qquad
\end{eqnarray}
Substituting these expressions into Eq.~(\ref{sigma v}), we therefore obtain
\begin{eqnarray}
    \langle \sigma v\rangle_{i}^{{\rm UV}} = \frac{A^{{\rm UV}}_{i}f_2^4}{8T^2}
\end{eqnarray}
and hence, using (\ref{n_i_thermal}), we have:
\begin{eqnarray}
    \Gamma^{\rm UV}_i = n_{i}\langle\sigma v\rangle_{i}^{\rm UV}= \frac{\zeta(3)}{8\pi^2}\xi_i A^{{\rm UV}}_{i}f_2^4\,T.\quad
\end{eqnarray}

\subsubsection{IR calculation}

In the IR (Einstein gravity) limit, the gravitational annihilation cross-section is instead given by 
\begin{equation}
    \sigma_{i}(s) = \frac{A^{{\rm IR}}_{i}s}{M_{pl}^4}
\end{equation} 
where\footnote{This can be obtained easily since, in our earlier calculation of $\sigma(s)$, the only difference is that the graviton propagator changes from $-2f_2^2\mathcal{P}^{(2)}_{\mu\nu\alpha\beta}/s^2$ in the UV limit to $4\mathcal{P}^{(2)}_{\mu\nu\alpha\beta}/sM_{pl}^2$ in the IR (Einstein gravity) limit.} $A^{{\rm IR}}_{i}=4A^{{\rm UV}}_{i}$ or, in other words
\begin{subequations}
  \begin{eqnarray}
    A_i^{\rm IR}&=&\frac{4}{3840\pi}\qquad(i={\rm spin}\;0) \\
    A_{i}^{\rm IR}&=&\frac{3}{3840\pi}\qquad(i={\rm spin}\,\frac{1}{2}) \\
    A_{i}^{\rm IR}&=&\frac{29}{3840\pi}\qquad(i={\rm spin}\;1).
  \end{eqnarray}
\end{subequations}

Thus, the integral in (\ref{sigma v}) becomes
\begin{eqnarray}
    &&\lim_{m\rightarrow 0}\int\limits_{4m^2}^{\infty}\sigma_{i}(s)(s-4m^2)\sqrt{s}\, K_1(\sqrt{s}/T)ds \nonumber\\ 
    &&=2\frac{A^{{\rm IR}}_{i}}{M_{pl}^4}T^7\int\limits_0^{\infty}x^6K_1(x)dx = 768\frac{A^{{\rm IR}}_{i}}{M_{pl}^4}T^7.\qquad
\end{eqnarray}
Substituting this and the prefactor expression (\ref{prefactor}) into Eq. (\ref{sigma v}) we therefore obtain
\begin{equation}
   \langle \sigma v\rangle_i^{{\rm IR}}=\frac{24A^{{\rm IR}}_{i}}{M_{pl}^4}T^2
\end{equation}
and hence, using (\ref{n_i_thermal}), we have:
\begin{eqnarray}
    \Gamma^{\rm IR}_i = n_{i}\langle \sigma v\rangle_{i}^{\rm IR}=\frac{24 \zeta(3)}{\pi^2}\xi_i A^{\rm IR}_{i}M_{pl}^{-4}T^5.\quad
\end{eqnarray}

\subsection{Final expression for the averaged gravitational annihilation rate $\overline{\Gamma}$}

Finally, we can collect the previous subsection's results to obtain the final expression for the average gravitational annihilation rate $\overline{\Gamma}$ as defined in Eq.~(\ref{def_Gamma_bar}), and conveniently re-expressed in Eq.~(\ref{expr_Gamma_bar}).  The result is
\begin{eqnarray}
\label{Gamma_bar_Appendix}
    \overline{\Gamma}
    =\left\{\begin{array}{ll}
    C_{{\rm UV}}f_{2}^{4}\,T &
    (T\gg f_{2}M_{pl}) \\
    C_{\,{\rm IR}\;}M_{pl}^{-4}\,T^{5} &
    (T\ll f_{2}M_{pl})
    \end{array}\right.
\end{eqnarray}
where 
\begin{subequations}
\label{C_Appendix}
  \begin{eqnarray}
      C_{\rm UV}&=&\frac{1393}{19365888 \pi^3}\approx
      2.8\times 10^{-6}, \\
      C_{\,{\rm IR}\;}&=&\;\;\,\frac{1393}{25216 \pi^{3}}\;\;\,\approx
      2.1\times 10^{-3}.
  \end{eqnarray}
\end{subequations}

\section{The shift $\Delta N_{eff}$ from a cosmic background of relic thermal gravitons}
\label{N_eff}

In this Appendix we show that, if the radiation era extends back to the bang (and, in particular, back to the Planck temperature $T\sim M_{pl}$, Einstein-gravity predicts that the universe today should be filled with a cosmic graviton background consisting of relic thermal gravitons (analogous to the cosmic microwave and neutrino backgrounds); and that this background, if it exists, would cause a specific shift in $N_{eff}$ (the number of ``effective neutrino species"), from $3.046$ to $3.046+\Delta N_{eff}$ where  $\Delta N_{eff}=(8/7)(43/441)^{4/3}$.  This shift might be detectable in forthcoming cosmology experiments like the Simons Observatory, and could be seen with high sensitivity at proposed experiments like CMB-S4 and CMB-HD.

If the hot radiation era {\it did} extend all the way back to the Planck temperature $T_{pl}$ then the number of relativistic species in the thermal bath included gravitons as well the Standard Model degrees of freedom. If, as we shall assume, the Standard Model includes right-handed neutrinos, one of them might be a stable particle and could be the dark matter. Unlike all the other Standard Model particles, it might never have thermalized. 

As the temperature dropped well below $T_{pl}$, the gravitons decoupled from the thermal bath.  Shortly after this decoupling, the number of effective degrees of freedom in equilibrium with the thermal bath was
\begin{equation}
  \label{gstar_sm}
  g_{\ast}^{(sm)}=\frac{7}{8}(2N_{1/2})+2N_{1}+N_{0}
  =\frac{441}{4},
\end{equation}
where $N_{0}=4$ is the number of real scalar fields in the Standard Model Higgs doublet, $N_{1}=8+3+1=12$ is the number of $SU(3)\times SU(2)\times U(1)$ gauge fields, and $N_{f}=(3\times 16)-1=47$ is the number of Weyl fermions in the Standard Model, excluding the dark matter ({\it i.e.},\ including a RH neutrino in each generation, but excluding one of those RH neutrinos -- the dark matter -- which never equilibrated with the thermal bath).\footnote{Note that if we didn't include RH neutrinos, we would instead have $N_{f}=3\times15=45$, and we would recover the familiar value $g_{\ast}=106.75$ for the minimal Standard Model.}

Later, when the temperature was around an MeV (just before neutrino decoupling), the remaining relativistic species in the thermal bath were: electrons (with two spin states), positrons (with two spin states), three neutrino flavours (each with one spin state), three anti-neutrino flavours (each with one spin state), and a photon (with two spin states), corresponding to an effective number of relativistic degrees of freedom
\begin{equation}
    g_{\ast}^{(\nu)}=\frac{7}{8}(2+2+3+3)+2=\frac{43}{4}.
\end{equation}
At this point, by entropy conservation, the temperature of the thermal bath (which is also the temperature $T_{\nu}$ inherited by the thermal bath of neutrinos when they decouple) is higher than the temperature $T_{gw}$ of the relic graviton background by a factor
\begin{equation}
  \frac{T_{\nu}}{T_{gw}}=\left(\frac{441}{43}\right)^{1/3}.
\end{equation}
Just after neutrino decoupling, the thermal bath included electrons, positrons and photons (with $g_{ast}=\frac{7}{8}(2+2)+2=\frac{11}{2}$), and a short time later the electrons and positrons annihilated, leaving just the photons (with $g_{\ast}=2$) as the only relativistic degrees of freedom in the thermal bath.  At this point, again by entropy conservation, the temperature of the thermal bath at (which is the temperature $T_{\gamma}$ inherited by the CMB) is higher than the temperature $T_{\nu}$ of the relic neutrino background by a factor
\begin{equation}
  \frac{T_{\gamma}}{T_{\nu}}=\left(\frac{11}{4}\right)^{1/3}.
\end{equation}

Now, we can compute the energy density of the relic (neutrino plus graviton) background relative to the energy density of the CMB:
\begin{equation}
  \frac{\rho_{\nu}+\rho_{gw}}{\rho_{\gamma}}
  =\frac{\frac{7}{8}(3+3)T_{\nu}^{4}+2T_{gw}^{4}}{2T_{\gamma}^{4}}.
\end{equation}

Plugging in our previous results, this becomes
\begin{equation}
  \frac{\rho_{\nu}+\rho_{gw}}{\rho_{\gamma}}
  =\frac{7}{8}\left(\frac{4}{11}\right)^{4/3}
  \left[3+\frac{8}{7}\left(\frac{43}{441}\right)^{4/3}\right]
\end{equation}

And finally, if we define the effective number $N_{eff}$ of relic neutrino species in the usual way \cite{ParticleDataGroup:2024cfk}
\begin{equation}
  \frac{\rho_{\nu}+\rho_{gw}}{\rho_{\gamma}}\equiv
  \frac{7}{8}\left(\frac{4}{11}\right)^{4/3}N_{eff}
\end{equation}
we see that the number of effective light neutrino species is shifted (from its standard value of $3.046$), to a new value
\begin{equation}
  N_{eff}=3.046+\Delta N_{eff}
\end{equation}
where
\begin{equation}
  \Delta N_{eff}=\frac{8}{7}\left(\frac{43}{441}\right)^{4/3}
  \approx 0.05129.
\end{equation}

This is compatible with the current observational limits \cite{ParticleDataGroup:2024cfk} from the CMB alone \cite{Planck:2018vyg}
\begin{equation} 
  N_{eff}=2.92^{+0.36}_{-0.37}\;\;
  (95\%{\rm CL}),
\end{equation}
or in combination with LSS \cite{Planck:2018vyg, DiValentino:2019dzu, ParticleDataGroup:2024cfk}
\begin{equation}
  N_{eff}=\left\{\begin{array}{ll} 
    2.99^{+0.34}_{-0.33} & (95\%{\rm CL}), \\
    2.85^{+0.23}_{-0.23} & (68\%{\rm CL}).
  \end{array}\right.
\end{equation}

For comparison, the Simons Observatory is projected to measure $N_{eff}$ with uncertainty $\sigma(N_{eff})=0.05$ \cite{SimonsObservatory:2018koc}, an experiment like CMB-S4 would achieve $\sigma(N_{eff})=0.02$ \cite{CMB-S4:2016ple}, and CMB-HD \cite{CMB-HD:2022bsz} would achieve $\sigma(N_{eff})=0.014$, a detection of the predicted shift at roughly the $4\sigma$ level.

\end{document}